\documentclass[11pt, journal, draftclsnofoot, onecolumn]{IEEEtran}
\IEEEoverridecommandlockouts
\usepackage[]{subfigure}
\usepackage{stfloats}
\usepackage{setspace}
\usepackage{cite, subfiles}
\usepackage{amsmath,amssymb,amsfonts, mathtools, amsthm}
\usepackage{stfloats}
\usepackage{algorithmic}
\usepackage{graphicx}
\usepackage{textcomp}
\usepackage{xcolor}
\usepackage{booktabs}
\usepackage{glossaries}
\usepackage{pgfplots}\pgfplotsset{compat=1.18}
\usepackage{algorithm}
\usepackage{tikz}
\usetikzlibrary{arrows,decorations.pathreplacing,patterns}
\usepackage{bbm}
\usepackage{bm}
\usepackage{multirow, tabularx}
\usepackage{dirtytalk}

\usepackage[normalem]{ulem}

\definecolor{palesilver}{rgb}{0.96, 0.92, 0.89}
\definecolor{antiquewhite}{rgb}{0.98, 0.92, 0.84}
\definecolor{lightcyan}{rgb}{0.94, 1.0, 1.0}
\definecolor{anti-flashwhite}{rgb}{0.95, 0.95, 0.96}
\definecolor{azure(web)(azuremist)}{rgb}{0.94, 1.0, 1.0}
\definecolor{beige}{rgb}{0.96, 0.96, 0.86}

\definecolor{redjigar}{rgb}{0.9, 0.01, 0.1}
\definecolor{chestnut}{rgb}{0.8, 0.36, 0.36}
\definecolor{airforceblue}{rgb}{0.36, 0.54, 0.66}
\definecolor{cadmiumorange}{rgb}{0.93, 0.53, 0.18}
\definecolor{bleudefrance}{rgb}{0.19, 0.55, 0.91}
\definecolor{carolinablue}{rgb}{0.6, 0.73, 0.89}
\definecolor{blue(ncs)}{rgb}{0.0, 0.53, 0.74}
\definecolor{dodgerblue}{rgb}{0.12, 0.56, 1.0}
\definecolor{cssgreen}{rgb}{0.0, 0.5, 0.0}
\definecolor{cadmiumgreen}{rgb}{0.0, 0.42, 0.24}
\definecolor{amaranth}{rgb}{0.9, 0.17, 0.31}
\definecolor{bluegray}{rgb}{0.4, 0.6, 0.8}
\definecolor{cadmiumgreen}{rgb}{0.0, 0.42, 0.24}
\definecolor{amethyst}{rgb}{0.6, 0.4, 0.8}
\definecolor{antiquebrass}{rgb}{0.8, 0.58, 0.46}

\def\BibTeX{{\rm B\kern-.05em{\sc i\kern-.025em b}\kern-.08em
    T\kern-.1667em\lower.7ex\hbox{E}\kern-.125emX}}

\newtheorem{conj}{Conjecture}

\UseRawInputEncoding

\title{Waveforms for Computing Over the Air}
\author{\IEEEauthorblockN{
Ana Perez-Neira\IEEEauthorrefmark{1}\IEEEauthorrefmark{2},
Marc Martinez-Gost\IEEEauthorrefmark{1}\IEEEauthorrefmark{2},
Alphan \c{S}ahin\IEEEauthorrefmark{3},
\\
Saeed Razavikia\IEEEauthorrefmark{4},
Carlo Fischione\IEEEauthorrefmark{4},
Kaibin Huang\IEEEauthorrefmark{5}}

\IEEEauthorblockA{
\IEEEauthorrefmark{1}Centre Tecnol\`{o}gic de Telecomunicacions de Catalunya, Spain\\
\IEEEauthorrefmark{2}Dept. of Signal Theory and Communications, Universitat Politècnica de Catalunya, Spain\\
\IEEEauthorrefmark{3}Dept. of Electrical Engineering, University of South Carolina, United States\\
\IEEEauthorrefmark{4}Dept. of Network and Systems Engineering, KTH University, Sweden\\
\IEEEauthorrefmark{5}Dept. of Electrical and Electronic Engineering, University of Hong Kong, Hong Kong}\\

Emails:
\{aperez, mmartinez\}@cttc.es,
asahin@mailbox.sc.edu\\
\{sraz, carlofi\}@kth.se, huangkb@hku.hk}

\begin{document}

\renewcommand{\labelenumii}{\arabic{enumi}.\arabic{enumii}}
\renewcommand{\labelenumiii}{\arabic{enumi}.\arabic{enumii}.\arabic{enumiii}}
\renewcommand{\labelenumiv}{\arabic{enumi}.\arabic{enumii}.\arabic{enumiii}.\arabic{enumiv}}
\def\timeSymbol{t}
\def\transmittedSignal[#1][#2]{\mathfrak{s}_{#1}(#2)}
\def\receivedSignal[#1][#2]{\mathfrak{r}_{#1}(#2)}
\def\superposedSignal[#1]{\mathfrak{r}(#1)}
\def\transmittedSignalPass[#1][#2]{\mathfrak{s}_{#1}'(#2)}
\def\receivedSignalPass[#1][#2]{\mathfrak{r}_{#1}'(#2)}

\def\fcarrierES{f_{\text{rx}}}
\def\fcarrierED[#1]{f_{\text{tx},#1}}
\def\phaseES{\theta_{\text{rx}}}
\def\phaseED[#1]{\theta_{\text{tx},#1}}
\def\indexED{k}

\def\numberOfActiveSubcarriers{M}
\def\indexTXsubcarrier{q}
\def\indexRXsubcarrier{\ell}
\def\indexRXsample{n}
\def\transmittedVectorEle[#1]{{{x}}_{#1}}

\def\ncfo[#1]{\eta_{\text{FO,}#1}}
\def\nto[#1]{\eta_{\text{TO,}#1}}
\def\nso{\eta_{\text{SO}}}
\def\cfo[#1]{\Delta f_{#1}}
\def\to[#1]{\Delta t_{\text{tx,}#1}}
\def\toComplete[#1]{\Delta t_{#1}}
\def\po[#1]{\Delta \theta_{#1}}
\def\so{\Delta t_{\text{rx}}}

\def\delaySymbol{\tau}
\def\delay[#1]{\delaySymbol_{#1}}
\def\gain[#1]{a_{#1}}
\def\gainComplex[#1]{h_{#1}}
\def\gainComplexCFR[#1]{H_{#1}}
\def\channelPass[#1]{\mathfrak{h}_{\indexED}'(#1)}
\def\channelBaseband[#1]{\mathfrak{h}_{\indexED}(#1)}
\def\pathIndex{w}
\def\numberOfPaths{W}
\def\deltaFunction[#1]{\delta(#1)}

\def\numberOfEdgeDevices{K}

\def\idftSize{N}
\def\symbolDuration{T_{\text{sym}}}
\def\cpDuration{T_{\text{cp}}}

\def\noiseVariance{\sigma_{\text{n}}^2}
\def\syncPoint{t_{\text{back-off}}}
\def\coefSync{\alpha}

\def\constante{{\rm e}}
\def\constantj{{\rm j}}
\def\complexNumbers{\mathbb{C}}
\def\realNumbers{\mathbb{R}}
\def\integers{\mathbb{Z}}
\def\energyMetric[#1]{e_{#1}}

\def\functionComp[#1]{f(#1)}
\def\parameter[#1]{s_{#1}}
\def\indexDigit{i}
\def\bit[#1]{b_{#1}}
\def\estimatorRatio[#1]{r_{#1}}

\def\affineEnc{g}
\def\affineDec{h}
\def\sequenceLength{N}

\def\precision{\epsilon}
\def\precisionF{p_0(\precision)}
\def\asmallNumber{\alpha}

\def\symbolVectorEle[#1]{{{s}}_{#1}}
\def\preProcessedVectorEle[#1]{{{p}}_{#1}}
\def\encodedVectorEle[#1]{{{c}}_{#1}}
\def\modulationVectorEle[#1]{{{m}}_{#1}}
\def\transmittedVectorEle[#1]{{{x}}_{#1}}
\def\superposedVectorEle[#1]{{{y}}_{#1}}
\def\modulationVectorSuperposedEle[#1]{{{\acute{m}}}_{#1}}
\def\demappedVectorEle[#1]{{{\acute{c}}}_{#1}}
\def\decodedVectorEle[#1]{{{\acute{p}}}{[#1]}}
\def\postProcessedVectorEle[#1]{{{z}}{[#1]}}
\def\noiseVectorEle[#1]{{{n}}_{#1}}
\def\complexNumbers{\mathbb{C}}
\def\aDomain{\mathbb{A}}
\def\compactInterval{\mathbb{E}}
\def\aCompactMetricSpace{\mathbb{S}}
\def\aMetricSpace{{\Gamma}}
\def\functionSpace[#1]{\mathcal{F}(#1)}
\def\realNumbers{\mathbb{R}}
\def\integers{\mathbb{Z}}
\def\constante{{\rm e}}
\def\constantj{{\rm j}}
\def\expectationOperator[#1][#2]{{\text{E}_{#2}}\left\{#1\right\}}
\def\uniformDistribution[#1][#2]{{\mathcal{U}_{[#1,#2]}}}
\def\traceOperator[#1]{{\textrm{tr}}\{#1\}}
\def\identityMatrix[#1]{\textbf{\textrm{I}}_{#1}}
\def\zeroVector[#1]{\textbf{\textrm{0}}_{#1}}
\def\exponentialIntegral[#1]{\textrm{Ei}(#1)}
\def\channelAtSubcarrier[#1]{h_{#1}}
\def\affineEnc{g}
\def\affineDec{h}
\def\weight[#1]{w_{#1}}
\def\indicatorFunction[#1]{\mathbb{I}\left[{#1}\right]}
\def\thresholdTCI{t}
\def\aFunction{H}
\def\normalizationCoef[#1]{\eta_{#1}}
\def\base{\beta}
\def\functionArbitrary[#1]{f_{#1}}
\def\functionArbitraryEstimate[#1]{\hat{f}_{#1}}
\def\preProcessingFunctionW[#1]{\psi_{#1}}
\def\postProcessingFunctionW[#1]{\varphi_{#1}}

\def\coefPoly[#1]{c_{#1}}
\def\indexPower{i}
\def\pForNorm{p}

\def\signNormal[#1]{\text{sign}\left(#1\right)}
\newcommand\mydots{\hbox to 1em{.\hss.\hss.}}

\newacronym{AirComp}{AirComp}{Over-the-air computation}
\newacronym{WSN}{WSN}{wireless sensor networks}
\newacronym{OFDM}{OFDM}{orthogonal frequency division multiplexing}
\newacronym{CDMA}{CDMA}{code division multiple access}
\newacronym{CSS}{CSS}{chirp spread spectrum}
\newacronym{CFO}{CFO}{carrier frequency offset}
\newacronym{AI}{AI}{artificial intelligence}
\newacronym{TO}{TO}{time offset}
\newacronym{PO}{PO}{phase offset}
\newacronym{LO}{LO}{local oscillator}
\newacronym{DFT}{DFT}{discrete Fourier transform}
\newacronym{CSI}{CSI}{channel state information}
\newacronym{FM}{FM}{frequency modulation}
\newacronym{PM}{PM}{phase modulation}
\newacronym{PAM}{PAM}{pulse amplitude modulation}
\newacronym{PCM}{PCM}{pulse code modulation}
\newacronym{PPM}{PPM}{pulse position modulation}
\newacronym{PWM}{PWM}{pulse width modulation}
\newacronym{SSB}{SSB}{single sideband}
\newacronym{DSB}{DSB}{double sideband}
\newacronym{QAM}{QAM}{quadrature amplitude modulations}
\newacronym{AWGN}{AWGN}{additive white Gaussian noise}
\newacronym{MSE}{MSE}{mean squared error}
\newacronym{SNR}{SNR}{signal-to-noise ratio}
\newacronym{MAC}{MAC}{multiple access channel}
\newacronym{TBMA}{TBMA}{type-based multiple access}
\newacronym{DA}{DA}{direct aggregation}
\newacronym{CSK}{CSK}{chirp-shift keying}
\newacronym{FSK}{FSK}{frequency-shift keying}
\newacronym{CPFSK}{CPFSK}{continuous phase frequency-shift keying}
\newacronym{PSK}{PSK}{phase-shift keying}
\newacronym{BPSK}{BPSK}{binary phase-shift keying}
\newacronym{QPSK}{QPSK}{quadrature phase-shift keying}
\newacronym{ASK}{ASK}{amplitude-shift keying}
\newacronym{APSK}{APSK}{amplitude phase-shift keying}
\newacronym{MSK}{MSK}{minimum shift keying}
\newacronym{ADC}{ADC}{analog-to-digital converter}
\newacronym{FEEL}{FEEL}{federated edge learning}
\newacronym{FDSS}{FDSS}{Frequency-Diversity Spread Spectrum}
\newacronym{FC}{FC}{fusion center}
\newacronym{SDR}{SDR}{software defined radio}
\newacronym{Log-FSK}{Log-FSK}{logarithmic FSK}
\newacronym{HDC}{HDC}{hyperdimensional computing}
\newacronym{DSSS}{DSSS}{direct sequence spread spectrum}
\newacronym{LoRa}{LoRa}{long range}
\newacronym{IoT}{IoT}{Internet of Things}
\newacronym{DCT}{DCT}{discrete cosine transform}
\newacronym{BER}{BER}{bit error rate}
\newacronym{PHY}{PHY}{physical layer}

\maketitle
\vspace{-50pt}
\section*{Abstract}
Over-the-air computation (AirComp) leverages the signal-superposition characteristic of wireless multiple access channels to perform mathematical computations. Initially introduced to enhance communication reliability in interference channels and wireless sensor networks, AirComp has more recently found applications in task-oriented communications, like wireless distributed learning and in wireless control systems. Its adoption aims to address latency challenges arising from an increased number of edge devices or \gls{IoT} devices accessing the constrained wireless spectrum. This paper is the first one to focus on the physical layer of these systems. We present a unified framework, specifically on the waveform and the signal processing aspects at the transmitter and receiver to meet the challenges that AirComp presents within the different contexts and use cases.

\newtheorem{theorem}{Theorem}
\newtheorem{acknowledgement}[theorem]{Acknowledgement}
\newtheorem{axiom}[theorem]{Axiom}
\newtheorem{case}[theorem]{Case}
\newtheorem{claim}[theorem]{Claim}
\newtheorem{conclusion}[theorem]{Conclusion}
\newtheorem{condition}[theorem]{Condition}
\newtheorem{conjecture}[theorem]{Conjecture}
\newtheorem{criterion}[theorem]{Criterion}
\newtheorem{definition}{Definition}
\newtheorem{exercise}[theorem]{Exercise}
\newtheorem{lemma}{Lemma}
\newtheorem{corollary}{Corollary}
\newtheorem{notation}[theorem]{Notation}
\newtheorem{problem}[theorem]{Problem}
\newtheorem{proposition}{Proposition}
\newtheorem{remark}{Remark}
\newtheorem{solution}[theorem]{Solution}
\newtheorem{summary}[theorem]{Summary}
\newtheorem{assumption}{Assumption}
\newtheorem{example}{\bf Example}
\newtheorem{probform}{\bf Problem}

\def\qed{$\Box$}
\def\QED{\mbox{\phantom{m}}\nolinebreak\hfill$\,\Box$}
\def\proof{\noindent{\emph{Proof:} }}
\def\poof{\noindent{\emph{Sketch of Proof:} }}
\def
\endproof{\hspace*{\fill}~\qed
\par
\endtrivlist\unskip}
\def\endproof{\hspace*{\fill}~\qed\par\endtrivlist\vskip3pt}

\def\E{\mathsf{E}}
\def\eps{\varepsilon}
\def\Lsp{{\boldsymbol L}}
\def\Bsp{{\boldsymbol B}}
\def\lsp{{\boldsymbol\ell}}
\def\Ltsp{{\Lsp^2}}
\def\Lpsp{{\Lsp^p}}
\def\Linsp{{\Lsp^{\infty}}}
\def\LtR{{\Lsp^2(\Rst)}}
\def\ltZ{{\lsp^2(\Zst)}}
\def\ltsp{{\lsp^2}}
\def\ltZt{{\lsp^2(\Zst^{2})}}
\def\ninN{{n{\in}\Nst}}
\def\oh{{\frac{1}{2}}}
\def\grass{{\cal G}}
\def\ord{{\cal O}}
\def\dist{{d_G}}
\def\conj#1{{\overline#1}}
\def\ntoinf{{n \rightarrow \infty }}
\def\toinf{{\rightarrow \infty }}
\def\tozero{{\rightarrow 0 }}
\def\trace{{\operatorname{trace}}}
\def\ord{{\cal O}}
\def\UU{{\cal U}}
\def\rank{{\operatorname{rank}}}
\def\acos{{\operatorname{acos}}}

\def\SINR{\mathsf{SINR}}
\def\SNR{\mathsf{SNR}}
\def\SIR{\mathsf{SIR}}
\def\tSIR{\widetilde{\mathsf{SIR}}}
\def\Ei{\mathsf{Ei}}
\def\l{\left}
\def\r{\right}
\def\({\left(}
\def\){\right)}
\def\lb{\left\{}
\def\rb{\right\}}

\setcounter{page}{1}

\newcommand{\eref}[1]{(\ref{#1})}
\newcommand{\fig}[1]{Fig.\ \ref{#1}}

\def\bydef{:=}
\def\ba{{\mathbf{a}}}
\def\bb{{\mathbf{b}}}
\def\bc{{\mathbf{c}}}
\def\bd{{\mathbf{d}}}
\def\bee{{\mathbf{e}}}
\def\bff{{\mathbf{f}}}
\def\bg{{\mathbf{g}}}
\def\bh{{\mathbf{h}}}
\def\bi{{\mathbf{i}}}
\def\bj{{\mathbf{j}}}
\def\bk{{\mathbf{k}}}
\def\bl{{\mathbf{l}}}
\def\bm{{\mathbf{m}}}
\def\bn{{\mathbf{n}}}
\def\bo{{\mathbf{o}}}
\def\bp{{\mathbf{p}}}
\def\bq{{\mathbf{q}}}
\def\br{{\mathbf{r}}}
\def\bs{{\mathbf{s}}}
\def\bt{{\mathbf{t}}}
\def\bu{{\mathbf{u}}}
\def\bv{{\mathbf{v}}}
\def\bw{{\mathbf{w}}}
\def\bx{{\mathbf{x}}}
\def\by{{\mathbf{y}}}
\def\bz{{\mathbf{z}}}
\def\b0{{\mathbf{0}}}

\def\bA{{\mathbf{A}}}
\def\bB{{\mathbf{B}}}
\def\bC{{\mathbf{C}}}
\def\bD{{\mathbf{D}}}
\def\bE{{\mathbf{E}}}
\def\bF{{\mathbf{F}}}
\def\bG{{\mathbf{G}}}
\def\bH{{\mathbf{H}}}
\def\bI{{\mathbf{I}}}
\def\bJ{{\mathbf{J}}}
\def\bK{{\mathbf{K}}}
\def\bL{{\mathbf{L}}}
\def\bM{{\mathbf{M}}}
\def\bN{{\mathbf{N}}}
\def\bO{{\mathbf{O}}}
\def\bP{{\mathbf{P}}}
\def\bQ{{\mathbf{Q}}}
\def\bR{{\mathbf{R}}}
\def\bS{{\mathbf{S}}}
\def\bT{{\mathbf{T}}}
\def\bU{{\mathbf{U}}}
\def\bV{{\mathbf{V}}}
\def\bW{{\mathbf{W}}}
\def\bX{{\mathbf{X}}}
\def\bY{{\mathbf{Y}}}
\def\bZ{{\mathbf{Z}}}

\def\mA{{\mathbb{A}}}
\def\mB{{\mathbb{B}}}
\def\mC{{\mathbb{C}}}
\def\mD{{\mathbb{D}}}
\def\mE{{\mathbb{E}}}
\def\mF{{\mathbb{F}}}
\def\mG{{\mathbb{G}}}
\def\mH{{\mathbb{H}}}
\def\mI{{\mathbb{I}}}
\def\mJ{{\mathbb{J}}}
\def\mK{{\mathbb{K}}}
\def\mL{{\mathbb{L}}}
\def\mM{{\mathbb{M}}}
\def\mN{{\mathbb{N}}}
\def\mO{{\mathbb{O}}}
\def\mP{{\mathbb{P}}}
\def\mQ{{\mathbb{Q}}}
\def\mR{{\mathbb{R}}}
\def\mS{{\mathbb{S}}}
\def\mT{{\mathbb{T}}}
\def\mU{{\mathbb{U}}}
\def\mV{{\mathbb{V}}}
\def\mW{{\mathbb{W}}}
\def\mX{{\mathbb{X}}}
\def\mY{{\mathbb{Y}}}
\def\mZ{{\mathbb{Z}}}

\def\cA{\mathcal{A}}
\def\cB{\mathcal{B}}
\def\cC{\mathcal{C}}
\def\cD{\mathcal{D}}
\def\cE{\mathcal{E}}
\def\cF{\mathcal{F}}
\def\cG{\mathcal{G}}
\def\cH{\mathcal{H}}
\def\cI{\mathcal{I}}
\def\cJ{\mathcal{J}}
\def\cK{\mathcal{K}}
\def\cL{\mathcal{L}}
\def\cM{\mathcal{M}}
\def\cN{\mathcal{N}}
\def\cO{\mathcal{O}}
\def\cP{\mathcal{P}}
\def\cQ{\mathcal{Q}}
\def\cR{\mathcal{R}}
\def\cS{\mathcal{S}}
\def\cT{\mathcal{T}}
\def\cU{\mathcal{U}}
\def\cV{\mathcal{V}}
\def\cW{\mathcal{W}}
\def\cX{\mathcal{X}}
\def\cY{\mathcal{Y}}
\def\cZ{\mathcal{Z}}
\def\cd{\mathcal{d}}
\def\Mt{M_{t}}
\def\Mr{M_{r}}
\def\O{\Omega_{M_{t}}}
\newcommand{\figref}[1]{{Fig.}~\ref{#1}}
\newcommand{\tabref}[1]{{Table}~\ref{#1}}

\newcommand{\var}{\mathsf{var}}
\newcommand{\fb}{\tx{fb}}
\newcommand{\nf}{\tx{nf}}
\newcommand{\BC}{\tx{(bc)}}
\newcommand{\MAC}{\tx{(mac)}}
\newcommand{\Pout}{p_{\mathsf{out}}}
\newcommand{\nnn}{\nn\\}
\newcommand{\FB}{\tx{FB}}
\newcommand{\TX}{\tx{TX}}
\newcommand{\RX}{\tx{RX}}
\renewcommand{\mod}{\tx{mod}}
\newcommand{\m}[1]{\mathbf{#1}}
\newcommand{\td}[1]{\tilde{#1}}
\newcommand{\sbf}[1]{\scriptsize{\textbf{#1}}}
\newcommand{\stxt}[1]{\scriptsize{\textrm{#1}}}
\newcommand{\suml}[2]{\sum\limits_{#1}^{#2}}
\newcommand{\sumlk}{\sum\limits_{k=0}^{K-1}}
\newcommand{\eqhsp}{\hspace{10 pt}}
\newcommand{\tx}[1]{\texttt{#1}}
\newcommand{\Hz}{\ \tx{Hz}}
\newcommand{\sinc}{\tx{sinc}}
\newcommand{\tr}{\mathrm{tr}}
\newcommand{\diag}{\mathrm{diag}}
\newcommand{\MAI}{\tx{MAI}}
\newcommand{\ISI}{\tx{ISI}}
\newcommand{\IBI}{\tx{IBI}}
\newcommand{\CN}{\tx{CN}}
\newcommand{\CP}{\tx{CP}}
\newcommand{\ZP}{\tx{ZP}}
\newcommand{\ZF}{\tx{ZF}}
\newcommand{\SP}{\tx{SP}}
\newcommand{\MMSE}{\tx{MMSE}}
\newcommand{\MINF}{\tx{MINF}}
\newcommand{\RC}{\tx{MP}}
\newcommand{\MBER}{\tx{MBER}}
\newcommand{\MSNR}{\tx{MSNR}}
\newcommand{\MCAP}{\tx{MCAP}}
\newcommand{\vol}{\tx{vol}}
\newcommand{\ah}{\hat{g}}
\newcommand{\tg}{\tilde{g}}
\newcommand{\teta}{\tilde{\eta}}
\newcommand{\heta}{\hat{\eta}}
\newcommand{\uh}{\m{\hat{s}}}
\newcommand{\eh}{\m{\hat{\eta}}}
\newcommand{\hv}{\m{h}}
\newcommand{\hh}{\m{\hat{h}}}
\newcommand{\Po}{P_{\mathrm{out}}}
\newcommand{\Poh}{\hat{P}_{\mathrm{out}}}
\newcommand{\Ph}{\hat{\gamma}}
\newcommand{\mat}[1]{\begin{matrix}#1\end{matrix}}
\newcommand{\ud}{^{\dagger}}
\newcommand{\C}{\mathcal{C}}
\newcommand{\nn}{\nonumber}
\newcommand{\nInf}{U\rightarrow \infty}

\section{Introduction}

\gls{AirComp} represents a groundbreaking approach in wireless communications that fundamentally redefines data aggregation. \gls{AirComp} enables a large number of distributed devices to combine their transmissions directly over the wireless channel, leveraging the waveform superposition property to compute mathematical nomographic functions \cite{goldenbaum13_aircomp}. Unlike the conventional transmit-then-compute approach, \gls{AirComp} performs computations directly in the air by exploiting signal superposition within the coherent \gls{MAC}. This process significantly reduces communication resources such as bandwidth and time. Instead of requiring each device to transmit data on a separate channel, \gls{AirComp} allows the unique combination of signals to be interpreted by the receiver without needing individual transmissions. In this way, for instance, 100 signals of 1 second in duration can be transmitted in 1 second instead of 100 seconds.

Originally developed for \gls{WSN}, \gls{AirComp} was revolutionary because it proved that uncoded transmission, specifically transmitting analog amplitude modulations of the observations, could be optimal under certain channel models \cite{nazer07_aircomp}. This technique leverages the structural alignment between the physical wireless channel and the sufficient statistic for estimating the source (i.e., the sum of the transmissions). Traditional separate source-channel strategies, where data is encoded into bits and then transmitted, introduce quantization noise and lead to a \gls{MSE} that scales logarithmically with the number of devices. In contrast, the analog transmission of \gls{AirComp} bypasses digital encoding, resulting in a lower estimation error that scales linearly.  This analog, in-channel computation model is especially well-suited for real-time applications, requiring minimum latency, as well as
large networks, where error accumulation is a concern \cite{gastpar06_liaison}.
This need has grown significantly with the expansion of Internet of Things (IoT) and distributed machine learning, where vast networks of sensors and devices transmit data to central points for processing. In traditional setups, each device would need to send data individually, resulting in delays, energy waste, and spectrum congestion. AirComp provides an efficient alternative by allowing simultaneous, uncoded transmissions that naturally combine in the air, achieving computations directly in the channel. Thus, \gls{AirComp} provides an efficient, scalable solution to support these applications by minimizing delays and reducing the computational load of individual devices.

Going further, as sensor networks generate increasingly vast amounts of data, \gls{AI} has emerged as a crucial framework for uncovering complex data relationships, expanding the range of wireless applications that require advanced computing beyond simple data fusion. This shift is commonly known as task-oriented applications \cite{Gunduz}, which are of paramount importance in sixth generation (6G) wireless networks. In 6G, communication systems will not only focus on data transmission but also on computing, indicating a significant paradigm shift. Future communication systems will prioritize enabling higher-level tasks over merely transmitting data, requiring redesign to meet specific application needs. For instance, reducing computational complexity and delays may be more important than minimizing the \gls{BER}. This transformation is exemplified by edge learning and distributed machine learning~\cite{Poor}, where the traditional wireless protocols fall short due to their inefficiency in handling the massive data volumes and individual transmissions from multiple devices. The reason is that they are oblivious to the ultimate goal, necessitating the development of innovative wireless communication methods to enhance spectrum efficiency and reduce latency. \gls{AirComp} is particularly promising for applications such as wireless data networks, wireless control systems, distributed localization, and wireless system-on-chip.

Let us note that \gls{AirComp} and NOMA (Non-Orthogonal Multiple Access) both enhance wireless network efficiency but address different needs. NOMA improves spectral efficiency by allowing multiple users to share the same frequency through power or code-domain techniques, ideal for high-density networks with varying user requirements \cite{islam2019noma}. In contrast, \gls{AirComp} enables real-time, in-channel computation by leveraging signal superposition, making it especially suitable for applications needing fast data aggregation. While NOMA focuses on user multiplexing, AirComp transforms the wireless channel into a computational tool, prioritizing low-latency data processing over user separation.

\gls{AirComp} is a disruptive multiple-access paradigm shift from compute-after-communicate to compute-when-communicate. To the authors' knowledge, in the literature on \gls{AirComp} (see \cite{Poor, sahin23_survey, wang22_foundations} as interesting overview papers), there is no work that discusses in detail the basics of the underlying signal processing methods at the transmitter and receiver for computation and waveform generation, and how they address the challenges in practical multiple access channels to achieve a successful transmission. Indeed, most of the existing works assume ideally aggregated signals at baseband, without paying attention to the required transmitted waveform or modulation to achieve this ideal model. Specifically, multipath fading, imperfect power control, and synchronization errors \cite{mish} can complicate or even invalidate the design of a reliable \gls{AirComp} scheme, where the receiver (e.g., fusion center) does not have access to the individual transmitted signals. \textit{How can multi-access be designed to ensure the most efficient collection of the transmitted data by a fusion center?} In this paper, we revisit the waveforms and the signal processing that can be used to jointly design the physical layer and multi-access wireless channel for the purpose of function computation. Both perfect and imperfect channel state knowledge are considered. 
We highlight the existence of many research gaps that have to be resolved in the field of signal processing to bring \gls{AirComp} to a practical implementation.
Currently, these evidences have been noticed by various authors under different perspectives and papers are proliferating in waveform design; thus, motivating this paper to present a systematic introduction of the fundamental signal processing techniques and guidelines to solve the physical layer for \gls{AirComp}. It is a well-defined, re-emerging area, and reasonably matured in some aspects, that deserves attention by the signal processing community as \gls{AirComp} opens new avenues for the researchers.

This paper is structured as follows: Section II states the signal processing problem of computing functions over multiple access channels, with a brief discussion on applications and metrics; Section III and IV respectively overview analog and digital modulations for \gls{AirComp} under perfect synchronization and perfect channel knowledge; Section V introduces the physical layer frame, which confers diversity and multiplexing capabilities to the system; Section VI presents the challenges that arise in \gls{AirComp} attending to the channel state information availability, and what techniques that can be implemented in this physical frame to resolve them; 
Section VII comments on the security concerns of \gls{AirComp}, and Section VIII concludes the paper and proposes future research directions.

\section{Computing functions over Multiple Access Channels}

\subsection{Applications}
\label{sec:applications}

In the era of big data, the growing interest in developing \gls{AirComp} schemes stems from the increasing number of applications that require computing a function over distributed data. In that case, \gls{AirComp} offers the potential to alleviate communication bottlenecks, such as delay, bandwidth limitations and power consumption, while also reducing the computational load on the receiver.  
By leveraging \gls{AirComp}, these computations can be executed more efficiently, and this paper explores various techniques for computing a wide range of functions, extending beyond the traditional sum or averaging operations.

The most prominent  application of \gls{AirComp} is in \gls{FEEL} \cite{Poor}, where the nodes collaboratively train a shared model by exchanging model parameters or gradients with a fusion center, such as an access point or a base station. Using \gls{AirComp}, all nodes can simultaneously transmit each model parameter and the receiver obtains a statistical average (e.g., arithmetic mean, median) of the models. The aggregated parameters are then distributed back to the devices for further updates until convergence.
For a comprehensive overview and the latest developments in FEEL, including discussions on architectures and learning algorithms, we direct the reader to \cite{liu2024recent}.
Another noteworthy application of \gls{AirComp} is on the MapReduce framework \cite{scholkopf2007mapreduce}. MapReduce is designed to process large datasets by decomposing complex functions into smaller sub-tasks that can be executed in parallel across distributed systems. In the context of data centers, \cite{wu16_stac} proposes the use of network-coded storage to address data erasures. Particularly, a weighted sum of the distributed data is used as a key to recover the data of a specific node, thereby enhancing the reliability and efficiency of data storage and retrieval in distributed systems.

Wireless sensor networks and control systems are other areas where \gls{AirComp} can be effectively applied for information aggregation. For instance, in \cite{song2011camera}, multiple cameras track a target to implement a distributed Kalman-consensus filter, where the sum of estimates from neighboring cameras is achieved through \gls{AirComp}. 
Similarly, \cite{liu2023pooling} demonstrates the implementation of a max-pooling operation in a distributed neural network via \gls{AirComp}.
In the context of vehicle platoons, \cite{lee2023platoon} describes an \gls{AirComp}-based policy for controlling the distance between cars, where statistical information such as maximum and minimum values play a crucial role.
Moreover, many standard signal processing techniques (e.g., singular-value decomposition, $k$-means, independent component analysis...) can be framed as weighted sum problems, making them suitable for \gls{AirComp}.
In \cite{sahin23_kmeans}, \gls{AirComp} is considered for the federated $k$-means clustering algorithm. 
Furthermore, \gls{AirComp} is increasingly utilized to interconnect numerous physically distributed in-memory computing cores, as seen in \gls{HDC}, which benefits from the concurrent transmissions enabled by \gls{AirComp} \cite{guirado23whype}.

\subsection{Problem Statement}\label{sec:Comp-MAC}

Consider a multiple access network comprising $K$ distributed nodes operating as data sources or transmitters, and a single node at the receiver side, referred to as receiver or fusion center. 
The goal of the network is to compute a function at the receiver side that represents the collective information from the distributed data. The transmitting nodes may be sensors collecting measurement readouts, or computing nodes, where data originates from local computations. 
Therefore, to maintain generality, we will not go into the model between the information source and the transmitting node. This is in contrast to other seminal works, such as \cite{gold1}, which deals solely with sensor networks and focuses on the channel between the source and the node. This broader approach ensures that our discussion is applicable to a variety of scenarios where the origin of the data is not the central focus, but rather the computation of the function from distributed inputs.

The $k$th node has a readout $s_k\in\mathbb{E}\triangleq[0,1]\in\mathbb{R}$ that remains constant over a period of time $T$. The network's goal to compute a function $f(s_1,\dots,s_K): \mathbb{E}^K\rightarrow\mathbb{R}$. We focus on symmetric functions, in which any permutation of the arguments of the function does not change the function output. Thus, only the data is relevant, not the origin nodes. Examples of such functions include the arithmetic mean, maximum, and majority voting, among other statistical functions. 
The $K$ signals are sent simultaneously, potentially sharing the same radio resources. This approach leverages the additive nature of the \gls{MAC} to compute the desired function $f$. At this stage, we assume a baseband model and $L$ orthogonal radio resources $l=1,\dots,L$ (i.e., time, frequency, or code), enabling the application of \gls{AirComp} to angular modulated signals, among other functionalities. Assuming perfect synchronization among the $K$ transmitters, the signal at the input of the receiver at the resource $l$ is 
\begin{equation}
    y_l = \sum_{k=1}^K h_{kl}\varphi_{kl}(s_k) + w_l,
    \label{eq:mac_baseband}
\end{equation}
where $h_{kl}$ is the channel coefficient between the sensor $k$ and the fusion center in the radio resource $l$; $\varphi_{kl}$ is the transmitted signal by the node $k$, which depends on the node data $k$ and acts on the same value $s_k$ for all $l$, and $w_l$ is the receiver noise.  Equation \eqref{eq:mac_baseband} is a high-level model abstraction that helps to introduce the problem. Later on we will complete \eqref{eq:mac_baseband} by introducing the modulations and processing that will allow a practical implementation. 

In the case of an ideal coherent \gls{MAC}, i.e., without fading channel and noise at the receiver, \eqref{eq:mac_baseband} reduces to
\begin{equation}
    y_l = \sum_{k=1}^K \varphi_{kl}(s_k),
    \label{eq:mac_baseband_ideal}
\end{equation}
which highlights the additive nature of the wireless MAC and allows to harness co-channel interference to compute a linear combination of functions of the distributed data.  An immediate consequence of this approach is a higher computation throughput, and, with it, a reduced latency or lower bandwidth requirements. For instance, to compute the arithmetic mean in an ideal \gls{MAC}, each transmitter transmits $\varphi_{kl}=s_k/K$ (assuming that the number of nodes in the computation is known at the transmitter), and the function is just the received symbol, i.e., $f=y_l$. In this case, notice that a single radio resource ($L=1$) is sufficient to compute $f$. 

The information-theoretic result of \cite{nazer07_aircomp} suggests that the superposition property of the wireless channel can be beneficially exploited if the MAC is matched in some mathematical sense to the function being computed. The approach is known as computation over the air or  \gls{AirComp}. Next, we generalize the model in \eqref{eq:mac_baseband_ideal} as


\begin{equation}
    \psi_l(y_l)=
    \psi_l\left( \sum_{k=1}^K \varphi_{kl}(s_k) \right),
    \label{eq:nomographic}
\end{equation}
where $\varphi_{kl}$ and $\psi_l$ are called preprocessing and postprocessing functions, respectively. Notice that both functions depend on the radio resource $l$. 
Encompassing preprocessing and postprocessing functions, the additive \gls{MAC} is transformed into an equivalent channel that aims to match the structure of the function to be computed. 
The multivariate functions that can be represented as in \eqref{eq:nomographic} are called \textit{nomographic} and are the fundamental building block for \gls{AirComp} systems \cite{goldenbaum15_lattice}. 
While this representation stems from an analytical function representation, researchers in the fields of signal processing and communications have noticed that such representation can be used for distributed computations: 1) Preprocessing functions can be computed at each transmitter in a distributed fashion, 2) Postprocessing functions are computed at the receiver side, and 3) The sum is performed by the \gls{MAC}. Any function computation approach that exploits the wireless \gls{MAC} via \eqref{eq:nomographic} is called an \gls{AirComp} scheme.

Finally, for $L>1$ the function of interest is obtained by appropriately combining the postprocessed measurements with an aggregation function $\Psi$:
\begin{equation}
    f(s_1,\dots,s_K)=
    \Psi\bigl(\psi_1(y_1),\dots,\psi_L(y_L)  \bigr).
    \label{eq:aggregation_function}
\end{equation}

Furthermore, computing a nomographic approximation enables reducing the number of radio resources. This is achieved by computing a nomographic function that approximates the original $f$ with respect to the precision $\epsilon$:
\begin{equation}
    \Big\|
    f-\Psi\left(\psi_1(y_1),\dots,\psi_L(y_L)\right)
    \Big\|_\infty\leq\epsilon.
    \label{eq:nomographic_approx}
\end{equation}

Table \ref{table:nomoFcns} displays the pre and postprocessing functions to compute several nomographic functions. In some cases the decomposition is exact, while in other cases, it is just an approximation according to \eqref{eq:nomographic_approx}.  For example, the geometric mean can be approximated with a preprocessing function $\ln(x+1/\precisionF)$ and a postprocessing function $\constante^{x/\numberOfEdgeDevices}$ for an arbitrary precision value $\precisionF>0$ as a function of the error $\epsilon$ in \eqref{eq:nomographic_approx}\cite{goldenbaum13_aircomp}.
{\color{black}
In \gls{AirComp}, however, the approximation is mandatory due to the existence of a wireless channel, even an ideal one. For instance, as we will show in Section \ref{sec:digital_mods}, working with digital schemes forces to represent $s_k$ in a digital base, such as bits. While \gls{AirComp} can be applied over each bit, different bits need to be transmitted orthogonally. Thus, each bit used to represent $s_k$ corresponds to a radio resource $l$. 
For this reason, while classical nomographic representations require no more than $L=1$ postprocessing functions to compute $f$ (see Table \ref{table:nomoFcns}), we may represent $f$ in another fashion that adapts to an appropriate transmission over the wireless channel. Thus, we highlight the contrast between the classical nomographic function representations as in Table \ref{table:nomoFcns}, and the nomographic decompositions that will be used in wireless communications.
}
Besides, in the case of a nonideal channel, we introduce $Q$ additional radio resources to counteract the channel impairments, such as noise or interference. They will be independent of the $L$ communication resources associated with postprocessing functions required to approximate $f$.

In a mathematical context, to describe any continuous and multivariate $f$, the Kolmorogov-Arnold theorem provides the existence of such representation using universal (i.e., independent of $f$) continuous preprocessing functions and multiple postprocessing functions. Particularly, $f$ can be represented as
\begin{equation}
    f(s_1,\dots,s_K)=
    \sum_{l=1}^{2K+1}\psi_l\left( \sum_{k=1}^K \varphi_{kl}(s_k) \right),
    \label{eq:kolmogorov}
\end{equation}
where the postprocessing function $\psi_l$ depends on $f$. 
According to the proposed model in \eqref{eq:aggregation_function}, the aggregation function $\Psi$ corresponds to the sum, and it requires $L=2K+1$ to compute the function.
The theorem proves the existence, not the construction of the corresponding preprocessing and postprocessing functions, and it does not identify $L$ with radio resources either.  In general, this theorem provides a fundamental guide, and we refer the curious reader to \cite{martinez23_enn}, where the authors use neural network architectures inspired by the Kolmorogov-Arnold theorem and propose Fourier series expansions to implement $\varphi_{kl}$.  


\begin{table}
	\centering
\caption{Example nomographic functions \cite{sahin23_survey}.}	
\resizebox{\textwidth*3/4-0.1in}{!}{
	\begin{tabular}{l|c|c|c|c}
		Description  & $\functionArbitrary[](\symbolVectorEle[1],\symbolVectorEle[2],\dots, \symbolVectorEle[\numberOfEdgeDevices])$ & $\postProcessingFunctionW[\indexED](x)$ & $\preProcessingFunctionW[](x)$   \\
		\hline\hline \begin{tabular}{@{}l@{}}Arithmetic mean\end{tabular} & $\displaystyle\frac{1}{\numberOfEdgeDevices}\sum_{\indexED=1}^\numberOfEdgeDevices\symbolVectorEle[\indexED]$ & $x$ & $\frac{x}{\numberOfEdgeDevices}$  \\
		\hline \begin{tabular}{@{}l@{}}Majority vote \end{tabular}& $\displaystyle\signNormal[{\sum_{\indexED=1}^{\numberOfEdgeDevices}\signNormal[{\symbolVectorEle[\indexED]}]}]$ & $\signNormal[x]$ & $\signNormal[x]$\\		
		\hline \begin{tabular}{@{}l@{}}$\pForNorm$-norm\end{tabular} & $\displaystyle\left({\sum_{\indexED=1}^\numberOfEdgeDevices|\symbolVectorEle[\indexED]|^{\pForNorm}}\right)^{1/\pForNorm}$ & $|x|^{\pForNorm}$ & $x^\frac{1}{\pForNorm}$ \\		
		\hline\hline \begin{tabular}{@{}l@{}} Approximation of \\ the geometric mean	\end{tabular} & $\displaystyle\left({\prod_{\indexED}\symbolVectorEle[\indexED]}\right)^{1/\numberOfEdgeDevices},\symbolVectorEle[\indexED]\ge0$ & $\ln\left(x+\frac{1}{\precisionF}\right)$ & $\constante^{\frac{x}{\numberOfEdgeDevices}}$ \\
		\hline\begin{tabular}{@{}l@{}} Approximation of \\  the maximum	\end{tabular} & $\displaystyle\max_{\indexED}\{\symbolVectorEle[\indexED]\},\symbolVectorEle[\indexED]\ge0$ & $x^{\precisionF}$ & $x^{\frac{1}{\precisionF}}$ \\
		\hline \begin{tabular}{@{}l@{}} Approximation of \\  the minimum\end{tabular} & $\displaystyle\min_{\indexED}\{\symbolVectorEle[\indexED]\},\symbolVectorEle[\indexED]\ge0$ & ${x^{-\precisionF}}$ & $x^{-\frac{1}{\precisionF}}$\\
		\hline
	\end{tabular}
}
\label{table:nomoFcns}
\end{table}

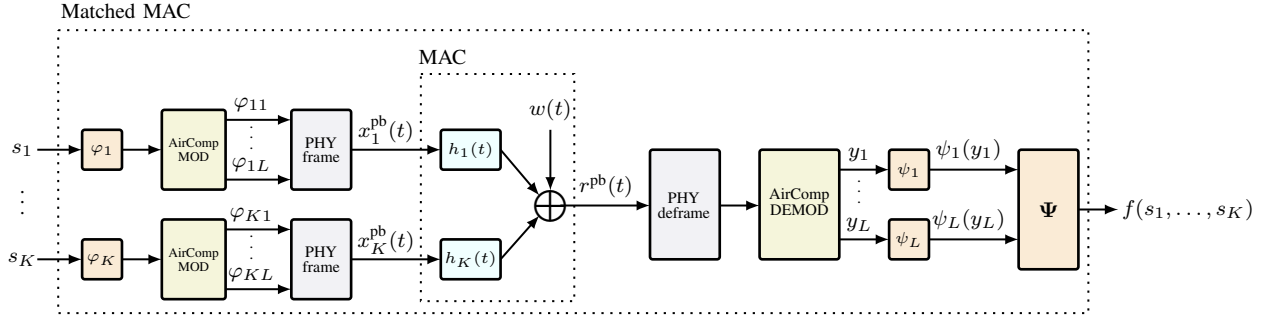
\begin{figure}
\centering
\tikzset{every picture/.style={line width=0.75pt}} 

\begin{tikzpicture}[x=0.75pt,y=0.75pt,yscale=-1,xscale=1]

\draw  [dash pattern={on 0.84pt off 2.51pt}] (28,20) -- (545,20) -- (545,162) -- (28,162) -- cycle ;

\draw  [dash pattern={on 0.84pt off 2.51pt}] (210,42) -- (287,42) -- (287,156) -- (210,156) -- cycle ;

\draw (28,5) node [anchor=north west][inner sep=0.75pt]  [font=\scriptsize] [align=left] {Matched MAC};
\draw (207.5,28) node [anchor=north west][inner sep=0.75pt]  [font=\scriptsize] [align=left] {MAC};

\draw (10,80) node  [font=\scriptsize]  {$s_{1}$};
\draw[-latex]    (17.5,80) -- (40,80) ;
	
\draw[rounded corners=1pt, line width=1pt, fill=antiquewhite]    
(40,70) -- (60,70) -- (60,90) -- (40,90) -- cycle ;
\draw (50,80) node  [font=\tiny]  {$\varphi_{1}$};
\draw (10,102.5) node [font=\scriptsize]  {$\vdots$};

\draw[-latex]    (60,80) -- (80,80) ;

\draw[rounded corners=1pt, line width=1pt, fill=beige]   (80,60) -- (112,60) -- (112,100) -- (80,100) -- cycle ;
\draw (96,80) node  [font=\tiny]  {$\substack{\text{AirComp}\\ \text{MOD}}$};

\draw (125,57.5) node  [font=\scriptsize]  {$\varphi_{11}$};
\draw[-latex]    (112,65) -- (145,65) ;
\draw (125,70) node [font=\tiny]  {$\vdots$};
\draw (125,87.5) node  [font=\scriptsize]  {$\varphi_{1L}$};
\draw[-latex]    (112,95) -- (145,95) ;

\draw[rounded corners=1pt, line width=1pt, fill=anti-flashwhite]      (145,60) -- (175,60) -- (175,100) -- (145,100) -- cycle ;
\draw (160,80) node  [font=\scriptsize]  {$\substack{\text{PHY}\\ \text{frame}}$};

\draw (192.5,70) node  [font=\scriptsize]  {$x_{1}^{\text{pb}}(t)$};
\draw[-latex]    (175,80) -- (220,80) ;

\draw[rounded corners=1pt, line width=1pt, fill=azure(web)(azuremist)]   (220,70) -- (250,70) -- (250,90) -- (220,90) -- cycle ;
\draw (235,80) node  [font=\tiny]  {$h_{1}(t)$};

\draw[-latex]    (250,80) -- (275-5.5,103) ;

\draw (275,108) node  [font=\large]  {$\bigoplus$};

\draw (10,135) node  [font=\scriptsize]  {$s_{K}$};
\draw[-latex]    (17.5,135) -- (40,135) ;

\draw [rounded corners=1pt, line width=1pt, fill=antiquewhite]   (40,125) -- (60,125) -- (60,145) -- (40,145) -- cycle ;
\draw (50,135) node  [font=\tiny]  {$\varphi_{K}$};
\draw[-latex]    (60,135) -- (80,135) ;

\draw[rounded corners=1pt, line width=1pt, fill=beige]   (80,115) -- (112,115) -- (112,155) -- (80,155) -- cycle ;
\draw (96,135) node  [font=\tiny]  {$\substack{\text{AirComp}\\ \text{MOD}}$};

\draw (125,112.5) node  [font=\scriptsize]  {$\varphi_{K1}$};
\draw[-latex]    (112,120) -- (145,120) ;
\draw (125,142.5) node  [font=\scriptsize]  {$\varphi_{KL}$};
\draw[-latex]    (112,150) -- (145,150) ;
\draw (125,125) node  [font=\tiny]  {$\vdots $};

\draw[rounded corners=1pt, line width=1pt, fill=anti-flashwhite]     (145,115) -- (175,115) -- (175,155) -- (145,155) -- cycle ;
\draw (160,135) node  [font=\scriptsize]  {$\substack{\text{PHY}\\ \text{frame}}$};

\draw (192.5,125) node  [font=\scriptsize]  {$x_{K}^{\text{pb}}(t)$};
\draw[-latex]    (175,135) -- (220,135);

\draw[rounded corners=1pt, line width=1pt, fill=azure(web)(azuremist)]    (220,125) -- (250,125) -- (250,145) -- (220,145) -- cycle ;
\draw (235,135) node  [font=\tiny]  {$h_{K}(t)$};

\draw[-latex]    (250,135) -- (275-5.5,113) ;

\draw[-latex]    (280,108) -- (325,108) ;

\draw (275,60) node  [font=\scriptsize]  {$w(t)$};
\draw[-latex]    (275,70) -- (275,100.5) ;
\draw (303,98) node  [font=\scriptsize]  {$r^{\text{pb}}(t)$};

\draw[rounded corners=1pt, line width=1pt, fill=anti-flashwhite]    (325,80) -- (360,80) -- (360,135) -- (325,135) -- cycle ;
\draw (342,108) node  [font=\scriptsize]  {$\substack{\text{PHY}\\ \text{deframe}}$};

\draw[-latex]    (360,108) -- (380,108) ;

\draw[rounded corners=1pt, line width=1pt, fill=beige]   (380,80) -- (420,80) -- (420,135) -- (380,135) -- cycle ;
\draw (400,106) node  [font=\scriptsize]  {$\substack{\text{AirComp}\\ \text{DEMOD}}$};

\draw[-latex]    (420,90) -- (445,90) ;

\draw[-latex]    (420,125) -- (445,125) ;

\draw (430,82.5) node  [font=\scriptsize]  {$y_{1}$};
\draw (430,97) node   [font=\tiny]  {$\vdots$};

\draw[rounded corners=1pt, line width=1pt, fill=antiquewhite]   (445,80) -- (465,80) -- (465,100) -- (445,100) -- cycle ;
\draw (455,90) node  [font=\tiny]  {$\psi_{1}$};
\draw (485,80) node  [font=\scriptsize]  {$\psi_{1}(y_{1})$};
\draw[-latex]    (465,90) -- (510,90) ;
\draw (430,117.5) node  [font=\scriptsize]  {$y_{L}$};
\draw[rounded corners=1pt, line width=1pt, fill=antiquewhite]   (445,115) -- (465,115) -- (465,135) -- (445,135) -- cycle ;
\draw (455,125) node  [font=\tiny]  {$\psi_{L}$};
\draw (485,115) node  [font=\scriptsize]  {$\psi_{L}(y_{L})$};
\draw[-latex]    (465,125) -- (510,125) ;

\draw[rounded corners=1pt, line width=1pt, fill=antiquewhite]   (510,80) -- (540,80) -- (540,140) -- (510,140) -- cycle ;
\draw (525,110) node  [font=\scriptsize]  {$\mathbf{\Psi}$};

\draw[-latex]    (540,110) -- (560,110) ;

\draw (595,110) node  [font=\scriptsize]  {$f(s_{1},\ldots ,s_{K})$};

\end{tikzpicture}
\caption{Multiple access network of $K$ distributed nodes implementing a modulation-based \gls{AirComp} scheme.}
\label{fig:AirComp_scheme}
\end{figure}

Fig. \ref{fig:AirComp_scheme} shows the system diagram to implement an \gls{AirComp} scheme over a \gls{MAC}. The \gls{AirComp} modulation (MOD) and demodulation (DEMOD) blocks are explicitly shown in the diagram. As discussed previously,
the modulator may generate up to $L$ preprocessing functions $\varphi_{kl}$ per transmitter, which depends on the function $f$ and the modulation of choice. The physical layer (PHY) frame encompasses the generation of a passband signal $x_{k}^{\text{pb}}(t)$ and manages the $Q$ radio resources to combat the channel impairments. At the receiver side, the signal is converted into a baseband signal by the corresponding deframing technique. The passband signal at the input of the receiver is
\begin{equation}
    r^{\text{pb}}(t) = \sum_{k=1}^K h_{k}(t)\circledast x_{k}^{\text{pb}}(t) + w(t),
    \label{eq:aircomp_analog}
\end{equation}
where $x_{k}^{\text{pb}}$ is the transmitted passband signal at node $k$, $\circledast$ corresponds to the convolutional operator, $h_k(t)$ is the passband channel between transmitter $k$ and the receiver, and $w(t)$ is the noise term at time $t$ and has power $\sigma_w^2$. 

Under no noise, perfect channel conditions and perfect channel knowledge, the demodulation should undo the modulation functions such that the sum of the transmitted signals is obtained; this differs from a conventional communication system, where the demodulation must separately recover each of the transmitted signals. 
In this way, we can take advantage of the aggregation in the \gls{MAC} in order to obtain an efficient system that does not require one orthogonal radio resource per transmitted signal. However, the receiver of an \gls{AirComp} scheme faces more difficult conditions than a non-\gls{AirComp} one {\color{black}because the receiver receives the signal after superposition. Hence, the AirComp receiver cannot simply compensate for the impact of distortion on the final output even if it has the perfect \gls{CSI} of each node (otherwise, it must deal with a highly overcomplete system of equations and recover the information transmitted from each transmitter, which defeats the purpose of AirComp). As discussed in Section~\ref{subsec:full CSI}, to address this issue, \gls{CSI} may be utilized at the transmitters with some precoders. However, we should note that the precoder used for typical communication scenarios (e.g., MIMO precoders), and the precoder for AirComp have different robustness to the distortions. For communication scenarios, when the corresponding precoder is “not” designed well, errors would not be fatal to the communication system. This is because all practical communication systems, like LTE, NR, and Wi-Fi, transmit some additional reference symbols that are also precoded (e.g., both data symbols and demodulation reference symbols (DMRS) in LTE and NR go through the same precoding procedure). Hence, even if the precoder is entirely random or suboptimal due to the erroneous \gls{CSI}, it would not be fatal to the communication as the equalization at the receiver based on the channel estimates over precoded reference symbols compensates for the composite response (the utilized precoder and the current \gls{CSI}). Such correction is not possible for AirComp, i.e., typical linear equalizers cannot be used, and distortions can be fatal. Also, a coherent AirComp requires phase synchronization among all nodes. From this aspect, the challenges for AirComp are similar to those of methods like interference alignment. The computation result for a coherent AirComp scheme will be sensitive to any node out of synch due to clock errors, residual carrier-frequency offset, calibration errors, and jittery time synchronization at both receiver and transmitters, as discussed in Section~\ref{subsec:SyncImpairments}. For this reason, \gls{AirComp} schemes widely vary and do not necessarily correspond to standard analog or digital techniques. When discussing these schemes for \gls{AirComp}, we will also discuss what demodulation procedures are required at the receiver.}

\subsection{Metrics}
\label{sec:metrics}
Given that the communication channel is not ideal, the receiver obtains $\hat{f}$, which is an estimate of the function $f$.
Since the goal of \gls{AirComp} is recovering a good estimate, the \gls{MSE} is the first metric to consider:
\begin{equation}
    \text{MSE}(\hat{f})\triangleq \mathbb{E}\{||f-\hat{f}||^2\},
    \label{eq:metric_mse}
\end{equation}
where the expectation is taken over the measurements and the channels. We also considers extensions of \eqref{eq:metric_mse}, such as the normalized \gls{MSE}. Alternatively, a more informative metric is \eqref{eq:metric_prob_out}. Given a maximum tolerable error $\varepsilon$, the outage probability measures the probability of exceeding the interval of confidence that $\varepsilon$ provides:
\begin{equation}
    P_{out}(\hat{f},\varepsilon)\triangleq
    \text{Prob}\left(\big|
    f-\hat{f}
    \big|\geq\varepsilon\right).
    \label{eq:metric_prob_out}
\end{equation}
When $f$ is discrete, an error only happens when $f\neq\hat{f}$, i.e., the symbol is misclassified. We define the computation error rate as
\begin{equation}
    P_{cer}(\hat{f})\triangleq
    \text{Prob}\left( \hat{f}\neq f \right).
    \label{eq:metric_prob_cer}
\end{equation}
Metrics \eqref{eq:metric_mse} and \eqref{eq:metric_prob_out} can be normalized, so that the difference between function values is normalized by the range of $f$.


Since the use of \gls{AirComp} reduces the need of orthogonal signaling, latency is largely reduced. 
While in this work we provide the most relevant metrics used for \gls{AirComp}, we refer the reader to \cite{sahin23_survey} for a extensive review of all the different metrics used in \gls{AirComp}.

It is important to note that depending on the application, there may be additional metrics that are not exclusively tied to \gls{AirComp}. Therefore, each application needs to analyze how its specific metrics relate to the fundamental \gls{AirComp} metrics outlined in this section.
For example, in task-oriented communications, the age of information measures the timeliness of data. Another example is the accuracy and convergence rate of \gls{FEEL}. However, these metrics not only vary depending on the application but are also currently under active research, so we exclude them from the scope of this paper.




\section{Analog modulations}
\label{sec:analog_mods}


While current communication systems exclusively utilize digital modulations, there is interest in reintroducing analog modulations driven by joint source-channel coding schemes. Additionally, it is essential to acknowledge that many successful systems still rely on analog modulations, such as commercial \gls{FM}, or \gls{PPM} and \gls{PWM} for instrumentation systems. These considerations are pertinent as future communication systems will integrate computation and sensing, requiring a holistic view of modulation techniques beyond their current applications in broadband or 5G communication systems.
Likewise, neural network systems empowered by the \gls{AI} native framework may impose complexity and latency constraints, prompting a reevaluation of analog systems. Overall, it is crucial to keep these aspects in mind and to reopen discussions that may have seemed closed within the field of communications.
Furthermore, the existing literature is significantly confusing using the terminology for the modulated waveforms. We will address and clarify these issues in the following sections.

\subsection{Amplitude modulations}

When multiple nodes transmit analog amplitude waveforms in a \gls{MAC}, the resulting waveform corresponds to an aggregation of the baseband signals. This occurs because the modulations are linear, meaning the signals simply add together.
We first introduce analog schemes that modulate information which is continuous in amplitude and time. These may be implemented when the node gathers information continuously, such as in sensing applications.
For \gls{DSB} modulation, the information $\varphi_k(s_k)$ modulates the amplitude of the carrier as  
\begin{equation}
    x_{k}^{\text{pb}}(t) = A_{k}\varphi_{k}(s_k)\cos\left(2\pi f_ct\right),
    \label{eq:linear_analog_mod}
\end{equation}
where $A_{k}$ controls the power of the transmitted signal by node $k$ and $f_c$ is the carrier frequency. The \gls{MAC} aggregation in \eqref{eq:aircomp_analog} particularized by \eqref{eq:linear_analog_mod} yields
\begin{equation}
    r^{\text{pb}}(t) = \sum_{k=1}^K h_{k}(t)\circledast A_{k}\varphi_{k}(s_k)\cos\left(2\pi f_ct\right) + w(t).
    \label{eq:aircomp_analog_fading}
\end{equation}
In the case of an ideal channel, i.e., $h_k(t)=\delta(t)$, and $A_k=A$,
\begin{equation}
    r^{\text{pb}}(t) = A\sum_{k=1}^K \varphi_{k}(s_k)\cos\left(2\pi f_ct\right) + w(t),
    \label{eq:aircomp_analog_fading_ideal}
\end{equation}
the amplitude of $r^{\text{pb}}(t)$ corresponds to the sum of the individual amplitudes. This linearity matches the ideal baseband MAC aggregation in \eqref{eq:mac_baseband_ideal}. In other words, the nomographic decompositions shown in Table~\ref{table:nomoFcns} can be applied straightforwardly, and amplitude modulations require no more than $L=1$ radio resource.

\textit{Demodulation}: According to \eqref{eq:aircomp_analog_fading_ideal}, the multiple access scheme used to implement \gls{AirComp} with amplitude modulations is \gls{DA}, this is, all users transmit simultaneously using the same radio source and the aggregation is performed seamlessly by the \gls{MAC}. It also implies that the \gls{AirComp} demodulation corresponds to the standard demodulation. For instance, in the case of \gls{DSB}, this corresponds either to synchronous or envelope detection \cite{carlson}. 
Afterward, the postprocessing $\psi$ is applied over the demodulated signal $y$, and the aggregation function $\Psi$ is redundant because $L=1$.
Besides, the \gls{SSB} modulation can be used to multiplex two signals simultaneously in the in-phase and quadrature components of the carrier.

As far as noise is concerned, \gls{DSB} with ideal synchronous detection has the same performance as analog baseband \gls{AirComp}, which is why linear modulations are implicitly considered in all works that solely develop \gls{AirComp} techniques in baseband \cite{zhu20_baa, goldenbaum13_robust, cao20_powercontrol, gold1}. However, the performance depends on the computed function because the postprocessing may alter the noise characteristics. For instance, the exponential postprocessing required to compute the geometric mean (see Table \ref{table:nomoFcns}), whenever it is possible to apply a logarithmic preprocessing, transforms the additive Gaussian noise into multiplicative and log-normal, which highly degrades the signal.



The high demand for power efficiency consolidated the development of analog pulse modulations. A time sampler obtains samples and modulates them into pulses.
Pulse-based modulations are also useful to optimally multiplex data in time. For instance, in \gls{FEEL}, each transmitter needs to send large amounts of parameters, which can be transmitted as a train of pulses. Alternatively, pulses are relevant in surveillance and \gls{IoT} applications, characterized by event-sampling \cite{premaratne2021event}.

Without loss of generality, we assume the transmission of a pulse every $T$. In analog \gls{PAM}, the transmitted waveform corresponding to a single pulse is
\begin{equation}
    x_{k}^{\text{pb}}(t) = A_{k}\varphi_{k}(s_k)p(t),
    \label{eq:pam_mod}
\end{equation}
where $p(t)$ is a finite-energy pulse of duration $T$. Since \gls{PAM} implements \gls{DA} as a multiple access scheme, the \gls{AirComp} demodulation follows the standard \gls{PAM} demodulation (e.g., low-pass filtering). 

{\color{black}
Because amplitude modulations preserve the linear aggregation of the baseband information, they seem the best choice for \gls{AirComp}. While this holds from a function approximation viewpoint,
amplitude modulations may not be ideal from the communication perspective. In practice, the transmitted power depends on the data and they are very sensitive to channel imperfections. For instance, they require tight \gls{CSI} to ensure the superposition required by \gls{AirComp}.} Overall, these aspects may push the power amplifier outside the linear region and incur high energy consumption. 


\subsection{Angular modulations}
Angular modulations place the information in the argument of the carrier. In the case of continuous-time information signals, these correspond to the classical frequency modulation \gls{FM} or \gls{PM}. 
Alternatively, the analog information can modulate the position or the width of transmitted pulses, thus receiving the name of \gls{PPM} and \gls{PWM}, respectively. 
The interest in angular modulations arises from their superior performance compared to amplitude modulations.
For instance, \gls{FM} requires moderately complex circuitry, while the \gls{SNR} can be increased arbitrarily consuming more bandwidth (i.e., the so-called power-bandwidth exchange) \cite{carlson}. 
However, due to their non-linear nature, the sum provided by the radio channel over the waveforms does not simply correspond the the sum of their arguments, this is, $\constante^{ja}+\constante^{jb}\neq \constante^{j(a+b)}$. Despite this, the lack of a multiple-access scheme integrating analog angular modulations for \gls{AirComp} does not preclude the use of digital angular modulations, as will be demonstrated in the next section.


\section{Digital modulations}
\label{sec:digital_mods}

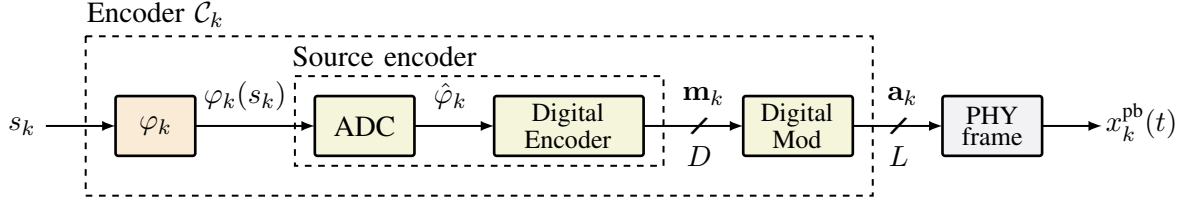
\begin{figure}
\centering
\tikzset{every picture/.style={line width=0.75pt}} 

\begin{tikzpicture}[x=0.75pt,y=0.75pt,yscale=-1]

\draw[rounded corners=1pt, line width=1pt, fill=antiquewhite]      (85,80) -- (125,80) -- (125,110) -- (85,110) -- cycle ;
\draw[-latex]    (50,95) -- (85,95) ;
\node  at (105,95) {$\varphi_k$} ;

\draw  [dash pattern={on 2.5pt off 2.5pt}] (175,70) -- (360,70) -- (360,115) -- (175,115) -- cycle ;
\draw [rounded corners=1pt, line width=1pt, fill=beige]    (185,80) -- (235,80) -- (235,110) -- (185,110) -- cycle ;

\draw[-latex]    (125,95) -- (185,95) ;
\draw[-latex]    (235,95) -- (275,95) ;
\draw [rounded corners=1pt, line width=1pt, fill=beige]  (275,80) -- (350,80) -- (350,110) -- (275,110) -- cycle ;
\draw[-latex]    (350,95) -- (400,95) ;
\draw[rounded corners=1pt, line width=1pt, fill=beige]   (400,80) -- (455,80) -- (455,110) -- (400,110) -- cycle ;
\draw  [dash pattern={on 2.5pt off 2.5pt}] (70,50) -- (465,50) -- (465,130) -- (70,130) -- cycle ;

\draw[-latex]    (455,95) -- (500,95) ;


\draw[rounded corners=1pt, line width=1pt, fill=anti-flashwhite]    (500,80) -- (550,80) -- (550,110) -- (500,110) -- cycle ;

\node at (525,95) {\Large $\substack{\text{PHY} \\ \text{frame}}$};

\draw[-latex]    (550,95) -- (580,95) ;

\draw (38,95) node    {$s_{k}$};

\draw (480,80) node    {$\mathbf{a}_{k}$};
\draw(478,95) node[rotate=-25]      {\textbf{/}};
\draw (478,110) node    {$L$};
\draw (600,93) node    {$x_{k}^{\text{pb}}(t)$};
\draw (150,80) node    {${\varphi}_{k}(s_k)$};
\node at (210,95) {ADC};

\draw (253,80) node    {$\hat{\varphi}_{k}$};
\node at (312,95) {\Large $\substack{\text{Digital} \\ \text{Encoder}}$};

\draw (380,80) node    {$\mathbf{m}_{k}$};
\draw (378,95) node[rotate=-25]      {\textbf{/}};
\draw (378,110) node    {$D$};

\node at (427,95) {\Large $\substack{\text{Digital} \\ \text{Mod}}$};

\draw (220,60) node   {Source encoder};
\draw (105,39) node   {Encoder $\mathcal{C}_k$};

\end{tikzpicture}
\caption{AirComp transmitter implemented with digital modulations.}
\label{fig:system_model_digital}
\end{figure}

Consider the digital transmission architecture in Fig. \ref{fig:system_model_digital}. 
First, the source encoder processes the input information through an \gls{ADC}, which discretizes the signal in both time and amplitude. We assume an $M$-level uniform quantizer, with $\hat\varphi_{k}\in[0,1,\dots,M-1]$ representing the quantized symbol. 
The next stage in the source encoder is the digital encoder, which produces a digital representation of the data $\mathbf{m}_k\in\mathbb{R}^D$. Historically, this has followed a binary representation, although we will show that alternative numeral systems may be useful for the purpose of \gls{AirComp}. For each quantized level, the encoder generates a vector of $D\leq L$ digits. This means that for each quantized symbol $\hat\varphi_{k}$, the source encoder produces $D$ components, which correspond to each of the preprocessing functions $\varphi_{kl}$. 

The vector $\mathbf{m}_k$ is fed to a digital modulator, which maps it to symbols $\mathbf{a_k}\in\mathbb{C}^L$ from a constellation. In traditional communication systems, the design of this constellation depends on factors such as the channel capacity and the desired \gls{BER}. 
However, for \gls{AirComp}, constellation design must align with specific \gls{AirComp} objectives, such as minimizing \gls{MSE}. 
Finally, each resulting symbol $\mathbf{a}_k$ is sent to the physical layer frame. This block is not only responsible for pulse shaping, power control, and up-conversion for passband, but also for building advanced modulation formats for multiplexing or diversity. Examples are multicarrier and spread spectrum modulations, as well as multi-antenna processing. Additionally, pilot sequences and other forms of redundancy, such as cyclic prefix, are also introduced in accordance with the physical layer protocol standards.

Note that in a practical wireless transmitter $\mathbf{m}_k$ undergoes channel encoding, which incorporates techniques to make the transmitted information robust to channel impairments. Fig. \ref{fig:system_model_digital} omits this detail to maintain the focus of this paper on other aspects, without delving into the specifics of channel encoder design.

In the following, we will introduce the existing digital modulations that can be implemented for \gls{AirComp}, along with the corresponding multiple access frameworks.

\subsection{Amplitude modulations}
As with analog amplitude modulations, digital amplitude modulations also align with the additive nature of the \gls{MAC}. Consider a digital unipolar \gls{PAM} with a normalized symbol amplitude of 1, such that $a_k\in[0,1,...,M-1]$. For this scheme to be effective in an \gls{AirComp} setup, there must be a linear mapping between the quantized levels $\hat\varphi_k$ and the \gls{PAM} symbols $a_k$ to preserve the additive nature of the \gls{MAC}. 
It is important to note that if the encoding introduces a nonlinearity (e.g., using a Gray code), the mapping from $\hat\varphi_k$ to $a_k$ is nonlinear and may cause overlaps between different aggregations. When the input variable includes both positive and negative values, digital polar \gls{PAM} should be used.
When the symbols are transmitted with a radio frequency pulse, it yields the \gls{ASK} modulation.
    \label{eq:digital_pam_mod}

\textit{Demodulation}: Notice that the receiver cannot use the same codebook for demodulation, since new symbols emerge due to the superposition of symbols. For instance, with $K=2$ transmitters, $a_1=1$ and $a_2=M-1$, the received symbol is $M$, which does not exist at the encoder side. Thus, the main drawback of digital amplitude modulations is that the codebook at the receiver depends on the number of users $K$.
Similar to analog modulations, two $M$-ary \gls{ASK} signals can be multiplexed in the phase and quadrature components of the waveform, which results in $M^2$-ary \gls{QAM}. 

Alternatively, the sum can be performed at the digit level as discussed in \cite{wu16_stac, guirado23whype}. For example, consider the quantized readout $\hat\varphi_k$ expressed in a $p$-ary system: $\hat\varphi_k=\sum_{d=0}^{D-1}p^dm_k^{(d)}$, where $m_k^{(d)}\in\mathbb{Z}_p$ is the $d$th digit of $\mathbf m_k\in\mathbb{Z}_p^D$. In binary case ($p=2$), where $m_k^{(d)}\in\{0,1\}$, the digital modulator maps each bit to a binary polar \gls{PAM} symbol: $a_{k}^{(d)}=1-2m_k^{(d)}$. Note that transmitting $D$ digits requires transmitting $D$ symbols multiplexed in $D$ orthogonal radio resources. Therefore, when \gls{AirComp} aggregates over digits, the original measurement $s_k$ is decomposed into $D$ preprocessed values $\varphi_{kl}= a_{k}^{(l)}$, with $D=L$, which are transmitted orthogonally over the \gls{MAC}. 

The received signal at resource $l$ after transmission through and ideal channel and demodulation is
\begin{equation}
    y_{l}=\sum_{k=1}^K a_k^{(l)}.
    \label{eq:bpsk_sum}
\end{equation}
The sum of the $l$th bit is equivalent to $\sum_{k=1}^K m_k^{(l)}=(K-y_l)/2$. Finally, as shown in \eqref{eq:binary_aggregation}, the arithmetic mean can be recovered by transforming the averaged $L$ bits to a decimal base. According to Fig. \ref{fig:AirComp_scheme}, equation \eqref{eq:binary_aggregation} encompasses both the postprocessing $\psi_l$ as well as the aggregation $\Psi$. The former corresponds to $\psi_l=2^l(K-y_l)/2K$, whereas $\Psi$ is a sum.
\begin{align}
    \frac{1}{K}
    \sum_{k=1}^K\hat{\varphi}_k=
    \frac{1}{K}
    \sum_{k=1}^K\sum_{l=1}^L 2^lm_k^{(l)}=
    \frac{1}{K}
    \sum_{l=1}^L 2^l\sum_{k=1}^K m_k^{(l)}=
    \sum_{l=1}^L 2^{l}\frac{K-y_l}{2K}.
    \label{eq:binary_aggregation}
\end{align}
In \cite{xie2023joint}, it is shown that the operation can lead to a carry, which requires the adoption of broader non-binary representations $(p>2)$. These aspects are revised in Subsection~\ref{subsec:Hybrid amplitude and angular modulations}, which introduces modulations with more degrees of freedom.

\subsection{Angular modulations}
\label{sec:digital_angular_mods}

As in analog angular modulations, digital angular modulations do not result in an aggregated signal that is linear with respect to the modulated information. However, modulations that provide an orthogonal waveforms, such as \gls{FSK} and \gls{PPM}, can be integrated into \gls{AirComp} using a multiple access scheme termed \gls{TBMA} \cite{mergen06_tbma}. This scheme leverages the fact that symmetric functions can be computed from the type (i.e., histogram) of the data. The core idea behind \gls{TBMA} is to establish a one-to-one mapping between $L$ different measurements and $L$ orthogonal radio resources. This means that in \gls{TBMA}, radio resources are allocated according to data, enabling superposition when users have the same measurement.

\textit{Demodulation}: After simultaneous transmission, the demodulation procedure corresponds to a bank of matched filters at the $L$ waveforms. The aggregated signal after demodulation (i.e., $y_1,\dots, y_L$ in Fig. \ref{fig:AirComp_scheme}) represents a histogram of the used radio resources, which equivalently corresponds to a histogram of the transmitted data. To compute $f$, the original authors of \gls{TBMA} propose an estimator for the received data distribution that asymptotically approaches the maximum likelihood estimator. This makes the \gls{TBMA} estimator consistent and efficient \cite{mergen06_tbma}. 
Note that, unlike traditional digital schemes, the \gls{TBMA} demodulation corresponds to a bank of matched filters, whose outputs (i.e., amplitudes) are used to estimate $f$.

\begin{figure}
\centering
\tikzset{every picture/.style={line width=0.75pt}} 

\begin{tikzpicture}[x=0.75pt,y=0.75pt,yscale=-1,xscale=1]

\draw (-10,80) node  [font=\scriptsize]  {$s_{1}$};
\draw (-10,110) node  [font=\scriptsize]  {$s_{2}$};
\draw (-10,140) node  [font=\scriptsize]  {$s_{3}$};
\draw (-10,190) node  [font=\scriptsize]  {$s_{10}$};

\draw (-10,160) node    {$\vdots$};

\draw[-latex]    (0,80) -- (30,80) ;
\draw[-latex]    (0,110) -- (30,110) ;
\draw[-latex]    (0,140) -- (30,140) ;
\draw[-latex]    (0,190) -- (30,190) ;


\draw[rounded corners=1pt, line width=1pt, fill=antiquewhite]   (30,70) -- (60,70) -- (60,90) -- (30,90) -- cycle ;
\draw (45,80) node  [font=\scriptsize]  {${\rm TX}_1$};

\draw[rounded corners=1pt, line width=1pt, fill=antiquewhite]    (30,100) -- (60,100) -- (60,120) -- (30,120) -- cycle ;
\draw (45,110) node  [font=\scriptsize]  {${\rm TX}_2$};

\draw[rounded corners=1pt, line width=1pt, fill=antiquewhite]    (30,130) -- (60,130) -- (60,150) -- (30,150) -- cycle ;
\draw (45,140) node  [font=\scriptsize]  {${\rm TX}_3$};

\draw[rounded corners=1pt, line width=1pt, fill=antiquewhite]    (30,180) -- (60,180) -- (60,200) -- (30,200) -- cycle ;
\draw (45,190) node  [font=\scriptsize]  {${\rm TX}_{10}$};

\draw   (60,80) -- (80,80) ;
\draw   (60,110) -- (80,110) ;
\draw   (60,140) -- (80,140) ;
\draw   (60,190) -- (80,190) ;

\draw   (80,80) -- (80,70) ;

\draw (80,110) -- (80,100);
\draw (80,140) -- (80,130);
\draw (80,190) -- (80,180);

\draw   (75,63) -- (80,70) -- (85,63) -- cycle ;
\draw   (75,93) -- (80,100) -- (85,93) -- cycle ;
\draw   (75,123) -- (80,130) -- (85,123) -- cycle ;
\draw   (75,173) -- (80,180) -- (85,173) -- cycle ;

\begin{scope}[shift={(20,70)}] 
 \draw[domain=74:132, smooth, variable=\x] plot ({\x},{10*sin(7*\x)});
\end{scope}
\begin{scope}[shift={(20,100)}] 
 \draw[domain=74:132, smooth, variable=\x] plot ({\x},{10*sin(12*\x)});
\end{scope}
\begin{scope}[shift={(20,130)}] 
 \draw[domain=74:132, smooth, variable=\x] plot ({\x},{10*sin(12*\x)});
\end{scope}

\begin{scope}[shift={(20,180)}] 
 \draw[domain=74:132, smooth, variable=\x] plot ({\x},{-10*sin(24*\x)});
\end{scope}

\draw[-latex]    (160,80) -- (220,115) ;
\draw[-latex]    (160,110) -- (220,120) ;
\draw[-latex]    (160,140) -- (220,125) ;
\draw[-latex]    (160,190) -- (220,135) ;

\draw    (230,130) -- (250,130) ;
\draw    (230,130) -- (230,120) ;

\draw   (225,113) -- (230,120) -- (235,113) -- cycle ;

\draw[rounded corners=1pt, line width=1pt, fill=antiquewhite]   (250,120) -- (280,120) -- (280,140) -- (250,140) -- cycle ;
\draw (265,130) node  [font=\scriptsize]  {${\rm RX}$};
\draw[-latex]    (280,130) -- (300,130) ;

\draw[rounded corners=1pt, line width=1pt, fill=beige]   (300,120) -- (330,120) -- (330,140) -- (300,140) -- cycle ;
\draw (315,130) node  [font=\scriptsize]  {${\rm DFT}$};
\draw[-latex]    (330,130) -- (350,130) ;

\draw[fill=beige]   (350,102) -- (460,102) -- (460,155) -- (350,155) -- cycle ;

\draw (360,135) -- (445,135) node[right] {\scriptsize $s$};
\draw (360,135) -- (360,110) ;
\draw (355,122) node  [font=\scriptsize, rotate=90]  {count};

\draw (368,145) node  [font=\scriptsize]  {$0$};
\draw (378,145) node  [font=\scriptsize]  {$1$};
\draw (388,145) node  [font=\scriptsize]  {$2$};
\draw (398,145) node  [font=\scriptsize]  {$3$};
\draw (408,145) node  [font=\scriptsize]  {$4$};
\draw (418,145) node  [font=\scriptsize]  {$5$};
\draw (428,145) node  [font=\scriptsize]  {$6$};
\draw (438,145) node  [font=\scriptsize]  {$7$};

\draw[fill=bluegray] (388-5,135) rectangle (388+5,128); 
\draw[fill=bluegray] (398-5,135) rectangle (398+5,114); 
\draw[fill=bluegray] (408-5,135) rectangle (408+5,114); 
\draw[fill=bluegray] (418-5,135) rectangle (418+5,121); 
\draw[fill=bluegray] (428-5,135) rectangle (428+5,128); 

\draw[-latex]    (460,130) -- (480,130) ;
\draw (515,130) node  [font=\scriptsize]  {$f(s_1,\ldots,s_{10})$};

\end{tikzpicture}
\caption{Example of a \gls{TBMA} implementation with \gls{FSK} in an ideal channel. We consider $K=10$ nodes with the following data: $(2,3,3,3,4,4,4,5,5,6)$. The system is designed with $M=8$ and each node transmits an \gls{FSK} carrier according to each measurement. At the receiver side, a bank of matched filters is implemented with the \gls{DFT}, and the result is a histogram of each waveform or measurement. The function is computed from the type.}
\label{fig:tbma_example}
\end{figure}
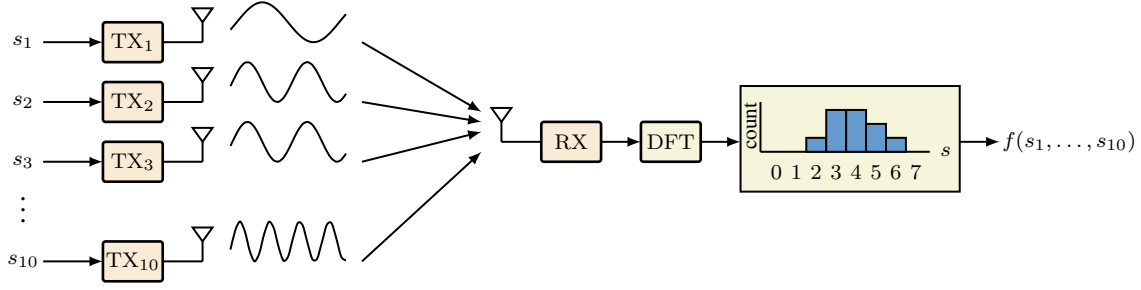

In Fig. \ref{fig:tbma_example} we show how to build a \gls{FSK}-based \gls{TBMA} system for \gls{AirComp}. To build a one-to-one mapping between the measurement and the radio resource spaces, the number of levels in the quantizer is set to $M=L$ and each $\mathbf m_k$ is mapped to a unique symbol $\mathbf{a}_k$. Under this framework, superposition over the \gls{MAC} happens for all nodes transmitting the same quantized readout $\hat\varphi_{k}$. Since \gls{TBMA} requires orthogonal signaling, it can implemented with $L$ orthogonal frequencies over $L$-ary \gls{FSK} \cite{martinez23_tbma}. 

The receiver deploys a bank of filters at the $L$ frequency bands, which can be effectively implemented using a \gls{DFT} of $L$ samples. The received signal after demodulation, $y_l$ corresponds to a noisy version of the number of nodes with the same measurement. In \gls{TBMA}, single-shot transmission by all users results in an increased bandwidth requirement. As an alternative to using the original \gls{TBMA} estimator, one can  compute the function from the received data. For example, the arithmetic mean can be computed as:
\begin{equation}
    \Psi(\psi_l(y_1),\dots,\psi_l(y_L))=\frac{1}{K}\sum_{l=1}^L ly_l,
    \label{eq:psi_tbma_mean}
\end{equation}
where $\psi_l=ly_l/K$ and $\Psi$ corresponds to the sum. Note that the aggregation does not necessarily correspond to the sum or even a linear function. For instance, the geometric mean can be computed as:
\begin{equation}
    \Psi(\psi_l(y_1),\dots,\psi_l(y_L))=\exp\left(\sum_{l=1}^{L} \frac{y_l}{K} \ln\left(l\right)
    \right).
    \label{eq:psi_tbma_geomean}
\end{equation}
Furthermore, \gls{TBMA} can be implemented at the bit level by assigning two orthogonal symbols (i.e., two radio resources) to each bit. The receiver recovers the type over each bit, which is used to compute the sum.
For signed variables, two's complement representation of integers can be used. This idea was exploited in \cite{xie2023joint} with a generic numeral system, and in \cite{sahin22_numerals} with a balanced number system to take the negative-valued parameters into account. 

\gls{TBMA} provides more detailed information to the receiver, though it comes with the cost of increased radio resource consumption. For instance, having access to the type and not just to a single aggregation by \gls{DA} enables the computation of multiple functions from a single transmission, as illustrated by \eqref{eq:psi_tbma_mean} and \eqref{eq:psi_tbma_geomean}. While this additional information does not necessarily lead to improved performance, it provides valuable benefits. For instance, it allows for the calculation of various functions and can also be useful for detecting anomalies or attackers within the system.

Furthermore, while linear analog modulations are framed as estimation problems, \gls{TBMA} depends on the function of interest. For instance, the arithmetic mean in \eqref{eq:psi_tbma_mean} corresponds to an estimation problem, while the maximum function corresponds to a detection. In the latter, the goal is to identify the signal associated with the largest radio resource. 
This distinction implies that the choice of the metric must align with the selected function and how it is implemented in terms of nomographic decomposition.
For estimation problems, one should consider the \gls{MSE}, as described in \eqref{eq:metric_mse}; for detection problems, the error rate, as given in \eqref{eq:metric_prob_cer}, is more relevant. It is important to notice that \gls{TBMA} is not the only method for \gls{AirComp} when using angular modulations, and Use Case $1$ illustrates an alternative approach.


\subsection*{Use case $1$: New angular modulations for \gls{DA}}
When two waveforms are multiplied instead of added, the product results in
\begin{equation}
    2\cos\left(2\pi f_1t\right)\cdot \cos\left(2\pi f_2t\right)=
    \cos\left(2\pi(f_1+f_2)t\right)+ \cos\left(2\pi(f_1-f_2)t\right).
    \label{eq:cos_prod}
\end{equation}
Due to the product, there is a term where the individual frequencies are added, which is the desired output of \gls{AirComp}. To achieve this product, we can transform the additive channel into a multiplicative one. As seen in Table~\ref{table:nomoFcns}, a sum turns into a product with a logarithmic preprocessing and an exponential postprocessing. If we apply the preprocessing step over the waveform, the result is
\begin{align}
    x_k(t) = A_k\log\left(\sqrt{\frac{2}{T}}\cos\left(\frac{\pi(2\hat\varphi_k+1)}{2T}t\right)+\alpha\right), 
    \label{eq:log_fsk}
\end{align}
and the exponential postprocessing is computed at the receiver side, in the demodulation. This waveform, referred to as \gls{Log-FSK}, leverages the nomographic representation to design the waveform \cite{martinez2024log}. With this, the matched \gls{MAC} allows to implement \gls{DA} with angular modulations. In light of \eqref{eq:cos_prod}, the demodulation procedure still requires a bank of filters and detecting the maximum frequency component, this is, where the \gls{AirComp} aggregation  happens. Since the amplitude is not relevant, \gls{Log-FSK} relies on pure digital demodulation.




\subsection*{Use case $2$: A hybrid analog and digital modulation for function approximation}
\label{sec:use_case_dct_fm}

Consider a single transmitter with data $s$, that needs to compute a function $f(s)$. While the \gls{DCT} has been used widely used to approximate and compress signals (see Fig. \ref{fig:sigmoid_DCT_FM}), its sinusoidal nature can be also exploited as a waveform. In \cite{gost23_dct}, an additional time index is appended to the inverse \gls{DCT} expression to transform it into a hybrid amplitude and angular modulation. The baseband transmitted signal is
\begin{equation}
    x(t) = A_c\sqrt{\frac{2}{N}}\sum_{i=1}^I F_i \cos\left(\frac{\pi i(2s+1)}{2T}t\right),
\label{eq:dct_mod}
\end{equation}
where $F_i, ~i=1,\dots,I$, are the \gls{DCT} coefficients that characterize function $f(s)$. The modulation is hybrid because it carries information in both the amplitude and the frequency of the signal. Thus, the waveform inherits the benefits from the \gls{DCT} representation. 
Fig. \ref{fig:MSE_DCT_FM} shows the \gls{MSE} to compute the sigmoid function with $I=3$. The \gls{DSB} carries $f(s)$ in the amplitude of the carrier, while \gls{FSK} corresponds to \eqref{eq:dct_mod} without the amplitudes, which are imposed at the receiver side.


\begin{figure}
\centering
\begin{minipage}{.45\textwidth}
  \centering
     \begin{tikzpicture} 
    \begin{axis}[
        xlabel={$s$},
        ylabel={$f(s)$},
        label style={font=\scriptsize},
        legend cell align={left},
        tick label style={font=\scriptsize} , 
        width=\textwidth,
        height=5cm,
        xmin=0, xmax=1,
        ymin=0, ymax=1,
       legend style={nodes={scale=0.55, transform shape}, at={(0.3,0.98)}}, 
        ymajorgrids=true,
        xmajorgrids=true,
        grid style=dashed,
        grid=both,
        grid style={solid, line width=.1pt, draw=gray!15},
        minor grid style={dotted, gray, line width=0.2pt},
    ]
    \addplot[
        color=bluegray,
        line width=1pt,
        mark size=2pt,
        ]
    table[x=x_axis,y=y]
    {images/Data/DCT_approx.dat};
    \addplot[dashed,
        color=cadmiumgreen,
        dashed, 
        line width=1pt,
        mark size=2pt,
        ]
    table[x=x_axis,y=y_3]
    {images/Data/DCT_approx.dat};
    \addplot[
        color=orange,
        dash pattern=on 1pt off 1pt on 4pt off 2pt, 
        line width=1pt,
        mark size=1pt,
        ]
    table[x=x_axis,y=y_7]
    {images/Data/DCT_approx.dat};
    \legend{$f(s)$, $I=3$, $I=7$};
    \end{axis}
\end{tikzpicture}
  \caption{Sigmoid function and its \gls{DCT} approximation with $I=3$ and $7$ coefficients.}
  \label{fig:sigmoid_DCT_FM}
\end{minipage}%
\hspace{20pt}
\begin{minipage}{.45\textwidth}
  \centering
   \begin{tikzpicture} 
    \begin{axis}[
        xlabel={$\rm SNR~(dB)$},
        ylabel={NMSE},
        label style={font=\scriptsize},
        legend cell align={left},
        tick label style={font=\scriptsize} , 
        width=\textwidth,
        height=5cm,
        xmin=-5, xmax=30,
         ymax=4.2e-4,
        ymode = log,
       legend style={nodes={scale=0.55, transform shape}, at={(0.98,0.98)}}, 
        ymajorgrids=true,
        xmajorgrids=true,
        grid style=dashed,
        grid=both,
        grid style={solid, line width=.1pt, draw=gray!15},
        minor grid style={dotted, gray, line width=0.2pt},
    ]
    \addplot[
        color=chestnut,
        line width=1pt,
        mark size=2pt,
        ]
    table[x=SNR,y=DCT_agnostic]
    {images/Data/NMSE_usecase2.dat};
    \addplot[dashed,
        color=bluegray,
        line width=1pt,
        mark size=2pt,
        ]
    table[x=SNR,y=DCT_nonagnostic]
    {images/Data/NMSE_usecase2.dat};
    \addplot[
        color=cadmiumgreen,
        dash pattern=on 1pt off 1pt on 4pt off 2pt,
        line width=1pt,
        mark size=1pt,
        ]
    table[x=SNR,y=SSB]
    {images/Data/NMSE_usecase2.dat};
    \legend{(21), DSB, FSK};
    \end{axis}
\end{tikzpicture}
  \caption{\gls{MSE} in an AWGN channel to approximate $f(s)$ with $I=3$.}
  \label{fig:MSE_DCT_FM}
  \end{minipage}
\end{figure}

\subsection{Hybrid amplitude and angular modulations}\label{subsec:Hybrid amplitude and angular modulations}



In Use Case 2 we have introduced a hybrid modulation to approximate a function in a communication channel. 
Hybrid modulations enable designing both the amplitude and the angle of the waveform. While standard \gls{APSK} modulations, such as \gls{QAM}, are not suitable for \gls{AirComp} due to destructive overlaps among the superimposed constellation points, these can be overcome by appropriately designing the digital modulator (see Use Case $3$). 

Following a more general approach, the goal is to uncover a proper choice for designing the digital modulator and preprocessing jointly so that \gls{DA} can be implemented for hybrid modulations. This joint design is performed by a set of encoders $\mathcal{C}_k$ and the decoder $\mathcal{D}$. If the optimum criterion is the minimum \gls{MSE}, it reads as:
\begin{equation}
    \mathcal{C}_1^*,\dots,\mathcal{C}_K^*, \mathcal{D}^* = \underset{\mathcal{C}_1,\dots,\mathcal{C}_K, \mathcal{D}}{\rm argmin} ~~\sum_{n=1}^N \Big\|{f}(s_1,\dots,s_K) - \Psi\bigl(\psi_1(y_1[n]),\dots,\psi_L(y_L[n])  \bigr)\Big\|_2^2~,
    \label{eq:main_goal_fading}
\end{equation}
where each encoder generates a constellation symbol from an input measurement: $a_{k}= \mathcal{C}_{k}(s_k)$ (see Fig.~\ref{fig:system_model_digital}). 
Since we are using digital modulations, let $R_f$ denote the number of all possible discrete values in the range of function $f$ and $s_{k}^{(i)}$ taking on one of $R_f$ possible values. Moreover, we define $f^{(i)}$ to be $i$th output of the function $f$ for a given set of quantized input values $s_{1}^{(i)}, s_{2}^{(i)}, \ldots, s_{K}^{(i)}$. Next, we use ${y}^{(i)}$ to present the constellation points corresponding to $i$th output value, i.e., ${y}^{(i)} = \sum_{k=1}^K a_k^{(i)}$, where $a_k^{(i)}\in\mathbb{C}~ \forall (i,k)$. Then, it is crucial that if $f^{(i)}$ differs from $f^{(j)}$ ($j$th output value), their respective constellation points ${y}^{(i)}$ and ${y}^{(j)}$ must also be different.  These conditions can be rewritten into the following feasibility optimization problem:
\begin{align}\label{eq:feasibility}
 &{\rm find} ~~a_1^{(i)}, \ldots,a_K^{(i)} ~~{\rm subject\, to}~~{\rm if~}
f^{(i)}\neq f^{(j)} \Rightarrow {y}^{(i)} \neq {y}^{(j)},~ 
 \forall (i,j), 
 ~|a_k^{(i)}|^2 \leq P_k, \quad \forall (i,k).
\end{align}
Generally, \eqref{eq:feasibility} is difficult to solve because it is highly non-convex and non-smooth. The authors in \cite{guirado2023whype} approach it in an intuitive fashion. 
Specifically, they investigate the bit-wise majority function $f={\rm sign}\big(\sum_{k=1}^K{\rm sign}(s_k)\big)$, by controlling the phase values in \gls{BPSK}. More specifically, for binary input values, $s_k^{(i)} = \{0,1\}$, the modulator becomes:
\begin{align}
a_k^{(i)} = A_k^{(i)}\left(1-2s_k^{(i)}\right), \quad \forall (i,k),
\end{align}
where $A_k^{(i)}\in\mathbb{C}$ has unit power. The imposition of the sign function constraint leads to a reduced number of feasible constellation diagrams, since $f$ is only either 0 or 1. Consequently, instead of solving the optimization in \eqref{eq:feasibility}, we can determine the transmitter phase shifts $A_k^{(i)}$ satisfying the constraints in \eqref{eq:feasibility} through an exhaustive search that yields the lowest \gls{BER}.

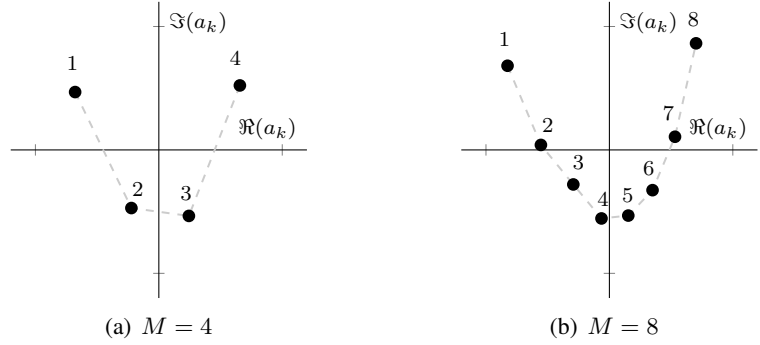
\begin{figure*}
\centering

%

\subfigure[$M = 4$]{
\label{fig:ConsQ2K4(b)}
   \begin{tikzpicture} 
    \begin{axis}[
        xlabel={\scriptsize \hspace{10pt} $\Re({a_k})$},
         x label style={
            at={(axis description cs:1,-0.1)}, 
        },
        ylabel={\scriptsize  $\Im({a_k})$},
        label style={font=\scriptsize},
        width=5.5cm,
        height=5.5cm,
        xmin=-6, xmax=6,
        ymin=-6, ymax=6,
        xticklabel=\empty,
        yticklabel=\empty,
        ticklabel style = {font=\tiny},
        major grid style={line width=1pt,draw=black},
        axis lines=center,
        axis line style={-},
        grid=none,
    ]
    \addplot[dashed,
        color=black!20,
        mark options={color=black, solid},
        mark=*,
        line width=0.75pt,
        mark size=2pt,
        ]
    table[x=X1,y=Y1]
    {images/Data/ProdModK4Q2.dat};4
     \node[] at (axis cs: -3.5pt,3.5pt) {\scriptsize $1$};
     \node[] at (axis cs: -0.85pt,-1.6pt) {\scriptsize $2$};
     \node[] at (axis cs: 1.1pt,-1.75pt) {\scriptsize $3$};
     \node[] at (axis cs: 3.1pt,3.7pt) {\scriptsize $4$};
    \end{axis}
  
\end{tikzpicture}
}\hspace{50pt}\subfigure[$M = 8$]{
\centering
\label{fig:ConsQ4K2(b)}
   \begin{tikzpicture} 
    \begin{axis}[
        xlabel={\scriptsize \hspace{10pt}$\Re({a_k})$},
        ylabel={\scriptsize  $\Im({a_k})$},
        label style={font=\scriptsize},
        width=5.5cm,
        height=5.5cm,
        xmin=-6, xmax=6,
        ymin=-6, ymax=6,
        xticklabel=\empty,
        yticklabel=\empty,
        ticklabel style = {font=\tiny},
        major grid style={line width=1pt,draw=black},
        axis lines=center,
        axis line style={-},
        grid=none,
    ]
    \addplot[dashed,
        mark=*,
        color=black!20,
        mark options={color=black, solid},
        line width=0.75pt,
        mark size=2pt,
        ]
    table[x=X1,y=Y1]
    {images/Data/ProdModK2Q3.dat};
     \node[] at (axis cs: -4.25pt,4.5pt) {\scriptsize $1$};
     \node[] at (axis cs: -2.5pt,1pt) {\scriptsize $2$};
     \node[] at (axis cs: -1.25pt,-0.5pt) {\scriptsize $3$};
     \node[] at (axis cs: -0.25pt,-2pt) {\scriptsize $4$};
     \node[] at (axis cs: 0.75pt,-1.85pt) {\scriptsize $5$};
     \node[] at (axis cs: 1.6pt,-0.75pt) {\scriptsize $6$};
     \node[] at (axis cs: 2.4pt,1.5pt) {\scriptsize $7$};
     \node[] at (axis cs: 3.4pt,5.3pt) {\scriptsize $8$};
    \end{axis}
\end{tikzpicture}}
  \caption{ Constellation diagram of the modulation vector for the product function $f =  \prod_{k=1}^Ks_k$ in different setups.} 
  \label{fig:ConsQ4K2}
\end{figure*}

On the other hand, the authors in \cite{razavikia23_channelcomp} study the general optimization in \eqref{eq:feasibility} for \gls{AirComp}. The non-convex and non-smooth constraints in \eqref{eq:feasibility} can be replaced with a smooth condition,
\begin{subequations}
\begin{align}
\nonumber
& \underset{\delta, a_1,\dots, a_K}{\text{maximize}} ~~~~ \delta \\
& \hspace{-20pt}\qquad\text{subject to}~~~
\Big|\sum_{k=1}^K a_k^{(i)}-\sum_{k=1}^K a_k^{(j)}\Big|^2  \geq \delta |f^{(i)}- f^{(j)}|^2,~\forall(i,j)\\
& ~~~~~~~~~~~~~~~
|a_k^{(i)}|^2 \leq P_k,~ \forall (i,k)
\end{align}
\label{eq:nonconvex_modulation}
\end{subequations}
where $P_k$ is the power budget for the constellation diagram at node $k$. Note that for any sufficiently small $\delta$, the solution to \eqref{eq:nonconvex_modulation} equals the solution to \eqref{eq:feasibility}. The optimization
problem in \eqref{eq:nonconvex_modulation} is a quadratic programming problem with non-convex constraints, known to be NP-hard~\cite{Sidir2006Physical}. Fortunately, this is a well-studied optimization that several approximation techniques have been developed to overcome the non-convex constraint, such as semi-definite relaxation~\cite{vandenberghe1996semidefinite} and Gaussian randomization
methods~\cite{luo2010semidefinite}.  Fig.~\ref{fig:ConsQ4K2} shows an example of the resulting constellations for computing product function $\prod_{k=1}^Ks_k$, where dots correspond to the constellation points used for transmission. Because the product function is symmetric, as defined in Section~\ref{sec:Comp-MAC}, the constellation points for both transmitter nodes are identical. When the modulation order increases from $M=4$ to $M=8$, the constellation diagram preserves the distribution of symbols along the same pattern.

\textit{Demodulation}: After solving the optimization problem in \eqref{eq:nonconvex_modulation}, the receiver needs to design the decision boundaries using the maximum likelihood estimator. Since the resulting constellation depends on the function, the corresponding de-mapper $\mathcal{D}(\cdot)$ changes with $f$ and $K$.




The main drawback of this approach is the complexity of solving such a non-convex optimization problem in \eqref{eq:feasibility}. To mitigate the design complexity, one effective approach is to confine the desired function to the sum function and limit the modulation optimization solely on the sum  function~\cite{razavikia2023sumcomp}. This allows the use of the same mapper $\mathcal{C}_k(\cdot) = \mathcal{C}(\cdot)$ across all nodes. Although not mandatory, the modulation mapper's capability to function as an additive map can simplify further the procedure. 

In this context, the linearity of nested lattice codes makes them suitable for accurately mapping inputs to constellation points. It is important to note that employing lattice codes for the transmitted symbols results in the superposed symbol at the receiver deviating from the input lattice field. Therefore, distinct from standard employed lattice codes in digital AirComp~\cite{goldenbaum15_lattice}, the focus shifts to the received codewords rather than the combination of transmitted codewords. This adjustment can potentially simplify the decoding process, as \cite{xie2023joint} shows, also enabling the straightforward adoption of low-density parity check channel coding (LDPC) schemes.

\subsection*{Use case $3$: \gls{DA} with angular modulations}

\begin{figure}
\centering
\scalebox{0.85}{
\begin{tikzpicture}[x=0.75pt,y=0.75pt,yscale=-1,xscale=1]

\draw  (-250,100) -- (-50,100) ;
\draw  (-150,0) -- (-150,200) ;

\draw (-30,100) node  {\footnotesize $\Re(a_k)$};
\draw (-150,-10) node  {\footnotesize $\Im(a_k)$};

\draw  (50,100) -- (250,100) ;
\draw  (150,0) -- (150,200) ;

\draw (270,100) node  {\footnotesize $\Re(a_k)$};
\draw (150,-10) node  {\footnotesize $\Im(a_k)$};

\draw[fill=white] (-150,100) node{}  circle  (4);
\draw[fill=white] (-75,100) node{}  circle   (4);
\draw[fill=white] (-225,100) node{}  circle  (4);

\draw (-140,90) node  {\footnotesize $0$};
\draw (-65,90) node  {\footnotesize $-1$};
\draw (-235,90) node  {\footnotesize $1$};

\draw[fill=white] (-150,175) node{}  circle  (4);

\draw[fill=white] (-75,175) node{}  circle   (4);

\draw[fill=white] (-225,175) node{}  circle  (4);

\draw (-140,185) node  {\footnotesize $-3$};
\draw (-65,185) node  {\footnotesize $-4$};
\draw (-235,185) node  {\footnotesize $-2$};

\draw[fill=white] (-150,25) node{}  circle  (4);
\draw[fill=white] (-75,25) node{}  circle   (4);
\draw[fill=white] (-225,25) node{}  circle  (4);

\draw (-140,15) node  {\footnotesize $3$};
\draw (-65,15) node  {\footnotesize $2$};
\draw (-235,15) node  {\footnotesize $4$};

\draw (-105,52.5) node  {\footnotesize $1$};
\draw (-195,52.5) node  {\footnotesize $2$};

\draw[fill=black] (-112.5,62.5) node{}  circle  (4);
\draw[fill=black] (-187.5,62.5) node{}  circle   (4);

\draw (-105,147.5) node  {\footnotesize $-2$};
\draw (-195,147.5) node  {\footnotesize $-1$};

\draw[fill=black] (-112.5,137.5) node{}  circle  (4);

\draw[fill=black] (-187.5,137.5) node{}  circle   (4);


\draw[fill=white] (150,100) node{}  circle  (4);
\draw[fill=white] (75,100) node{}  circle   (4);
\draw[fill=white] (225,100) node{}  circle  (4);

\definecolor{chestnut}{rgb}{0.9, 0.2, 0.25}
\draw (187,90) node  {\footnotesize {\color{chestnut}$3\neq -3$}};
\draw (65,90) node  {\footnotesize $1$};
\draw (235,90) node  {\footnotesize $-1$};

\draw[fill=white] (150,175) node{}  circle  (4);

\draw[fill=white] (75,175) node{}  circle   (4);

\draw[fill=white] (225,175) node{}  circle  (4);

\definecolor{cadmiumorange}{rgb}{1, 0.6, 0.1}
\draw (140,185) node  {\footnotesize {\color{cadmiumorange} $0$}};
\draw (65,185) node  {\footnotesize $4$};
\draw (235,185) node  {\footnotesize $-4$};

\draw[fill=white] (150,25) node{}  circle  (4);
\draw[fill=white] (75,25) node{}  circle   (4);
\draw[fill=white] (225,25) node{}  circle  (4);

\draw (140,15) node  {\footnotesize {\color{cadmiumorange} $0$}};
\draw (65,15) node  {\footnotesize $-2$};
\draw (235,15) node  {\footnotesize $2$};

\draw (105,52.5) node  {\footnotesize $-1$};
\draw (195,52.5) node  {\footnotesize $1$};

\draw[-latex, chestnut, line width=0.75pt] (117,67.5) -- (145,95);
\draw[-latex, dashed, chestnut, line width=0.75pt] (183,67.5) -- (155,95);

\draw[-latex, dashed, chestnut, line width=0.75pt] (117,132.5) -- (145,105);
\draw[-latex, chestnut, line width=0.75pt] (183,132.5) -- (155,105);

\draw[-latex, cadmiumorange, line width=0.75pt] (117,57.5) -- (145,30);
\draw[-latex, cadmiumorange, line width=0.75pt] (183,57.5) -- (155,30);

\draw[-latex, dashed, cadmiumorange, line width=0.75pt] (117,142.5) -- (145,170);
\draw[-latex, dashed, cadmiumorange, line width=0.75pt] (183,142.5) -- (155,170);

\draw[fill=black] (112.5,62.5) node{}  circle  (4);
\draw[fill=black] (187.5,62.5) node{}  circle   (4);

\draw (105,147.5) node  {\footnotesize $2$};
\draw (195,147.5) node  {\footnotesize $-2$};

\draw[fill=black] (112.5,137.5) node{}  circle  (4);

\draw[fill=black] (187.5,137.5) node{}  circle   (4);

\end{tikzpicture}
}

\caption{Constellation diagram showcasing the transmission and reception of \gls{QPSK} signals by two nodes ($K=2$) to compute the sum. The black dots correspond to the constellation used for transmission, while the white dots correspond to the constellation at the receiver side, i.e., after \gls{AirComp} superposition. We assume $\varphi_k=\{-2,1,1,2\}$ and propose two different modulation mappings to symbols. Red and orange arrows indicate destructive overlaps and redundant symbols, respectively.}
\label{fig:Overlaps}
\end{figure}
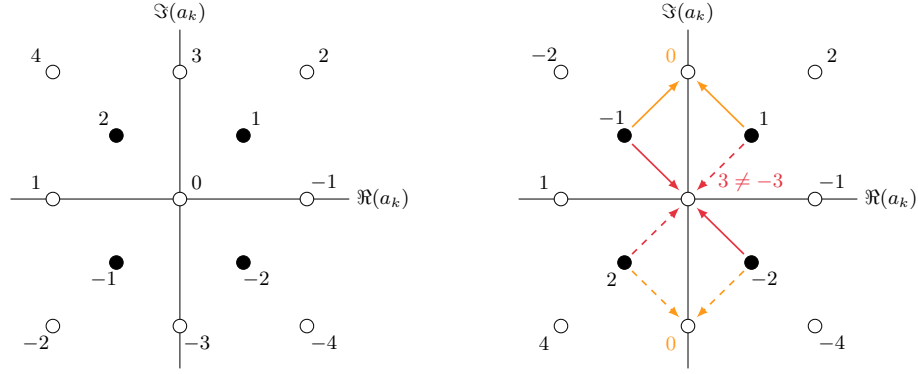

Implementing \gls{DA} using existing digital modulation schemes presents significant challenges, mainly because the aggregation of symbols generates overlaps that correspond to different function values, leading to ambiguity at the receiver.
While most digital modulations struggle with this issue, \gls{BPSK}, which corresponds to a binary polar \gls{PAM}, as in \eqref{eq:bpsk_sum}, is an exception and can be used effectively for DA.
The extension to higher-order modulations is cumbersome but not impossible. Fig. \ref{fig:Overlaps} shows the use of \gls{QPSK} modulations for $K=2$ and $\hat\varphi_k=\{-2,-1,1,2\}$. In the left figure, an appropriate mapping of $\hat\varphi_k$ to the constellation symbols $a_k$ makes it possible to implement \gls{AirComp}. While overlaps occur, these correspond to the same function value, which does not generate any ambiguity on the receiver side. Conversely, in the right figure, we show that not every modulation mapping is possible. Swapping two symbols in the first constellation results in destructive overlaps, making it impossible for the receiver to distinguish between different function values (i.e., $3$ and $-3$). Additionally, this mapping results in redundant symbols (e.g., symbol $0$), which further complicates the demodulation process. 

Extending this encoding approach to networks with more than $K=2$ nodes introduces even greater challenges. The number of overlaps and redundant symbols tends to increase with 
$K$, making it increasingly difficult to maintain the reliability and efficiency of the DA process in larger networks. To simplify DA in larger networks, the SumComp method in \cite{razavikia2023sumcomp} designs the appropriate mapping of the constellation symbols $a_k$ in an optimization free method for computing the nomographic function over-the-air.  However, the SumComp mapping is restricted to a modulation family involving QAM, hexagonal QAM, and PAM.

\definecolor{bluegray}{rgb}{0.4, 0.6, 0.8}
\definecolor{bazaar}{rgb}{0.6, 0.47, 0.48}
\definecolor{darklavender}{rgb}{0.45, 0.31, 0.59}
\definecolor{cadmiumgreen}{rgb}{0.0, 0.42, 0.24}

\begin{figure*}
\centering
\subfigure[Low-density scenario ($K=10$).]{
 \label{fig:Sim_low}
    \begin{tikzpicture} 
    \begin{axis}[
        xlabel={$\rm SNR~(dB)$},
        ylabel={NMSE},
        label style={font=\scriptsize},
        legend cell align={left},
        tick label style={font=\scriptsize} , 
        width=0.48\textwidth,
        height=6cm,
        xmin=-15, xmax=19,
        ymin=2e-4, ymax=5,
        ymode = log,
       legend style={nodes={scale=0.45, transform shape}, at={(0.98,0.98)}}, 
        ymajorgrids=true,
        xmajorgrids=true,
        grid style=dashed,
        grid=both,
        grid style={solid, line width=.1pt, draw=gray!15},
        minor grid style={dotted, gray, line width=0.2pt},
    ]
    \addplot[
        color=darklavender,
        mark=o,
        line width=1pt,
        mark size=2pt,
        ]
    table[x=SNR,y=ChannelSumA]
    {images/Data/NMSEerrorwave.dat};
    \addplot[
        color=bazaar,
        mark=o,
        line width=1pt,
        mark size=2pt,
        ]
    table[x=SNR,y=ChannelCompA]
    {images/Data/NMSEerrorwave.dat};
     \addplot[
        color=bluegray,
        mark=o,
        line width=1pt,
        mark size=2pt,
        ]
    table[x=SNR,y=OATA]
    {images/Data/NMSEerrorwave.dat};
    \addplot[
        color=darklavender,
        mark=square,
        line width=1pt,
        mark size=2pt,
        ]
    table[x=SNR,y=ChannelSumG]
    {images/Data/NMSEerrorwave.dat};
    \addplot[
        color=bazaar,
        mark=square,
        line width=1pt,
        mark size=2pt,
        ]
    table[x=SNR,y=ChannelCompG]
    {images/Data/NMSEerrorwave.dat};
     \addplot[
        color=bluegray,
         mark=square,
        line width=1pt,
        mark size=2pt,
        ]
    table[x=SNR,y=OATG]
    {images/Data/NMSEerrorwave.dat};
    \legend{SumComp $\sum$ $(M=64)$, ChannelComp~$\sum$ $(M=64)$, Analog \gls{DA}~$\sum$, SumComp $\prod$ $(M=8)$, ChannelComp~$\prod$ $(M=8)$, Analog \gls{DA}~ $\prod$};
    \end{axis}
\end{tikzpicture}}\subfigure[High-density scenario ($K=100$).]{
 \label{fig:Sim_high}
    \begin{tikzpicture} 
    \begin{axis}[
        xlabel={$\rm SNR~(dB)$},
        ylabel={NMSE},
        label style={font=\scriptsize},
        legend cell align={left},
        tick label style={font=\scriptsize} , 
        width=0.48\textwidth,
        height=6cm,
        xmin=-15, xmax=19,
        ymin=1e-5, ymax=5,
        ymode = log,
       legend style={nodes={scale=0.55, transform shape}, at={(0.98,0.98)}}, 
        ymajorgrids=true,
        xmajorgrids=true,
        grid style=dashed,
        grid=both,
        grid style={solid, line width=.1pt, draw=gray!15},
        minor grid style={dotted, gray, line width=0.2pt},
    ]
    \addplot[
        color=cadmiumgreen,
        mark=o,
        line width=1pt,
        mark size=2pt,
        ]
    table[x=SNR,y=TBMA_Sum]
    {images/Data/NMSE_tbma_lin.dat};
    \addplot[
        color=bluegray,
        mark=o,
        line width=1pt,
        mark size=2pt,
        ]
    table[x=SNR,y=Analog_DA_Sum]
    {images/Data/NMSE_tbma_lin.dat};
    \addplot[
        color=cadmiumgreen,
        mark=square,
        line width=1pt,
        mark size=2pt,
        ]
    table[x=SNR,y=TBMA_Prod]
    {images/Data/NMSE_tbma_lin.dat};
     \addplot[
        color=bluegray,
         mark=square,
        line width=1pt,
        mark size=2pt,
        ]
    table[x=SNR,y=Analog_DA_Prod]
    {images/Data/NMSE_tbma_lin.dat};
    \legend{TBMA $\sum$ $(M=64)$, Analog \gls{DA}~ $\sum$, TBMA $\prod$ $(M=64)$, Analog \gls{DA}~ $\prod$};
    \end{axis}
\end{tikzpicture}}
  \caption{Performance comparison of different \gls{AirComp} schemes for the arithmetic ($\Sigma$) and geometric mean ($\Pi$) functions. In {\color{black} Fig.~\ref{fig:Sim_low}}, we compare \gls{DA} schemes in the analog and digital domain. In {\color{black} Fig.~\ref{fig:Sim_high}}, we compare \gls{TBMA} to analog \gls{DA}. The input values are generated  $s_k \sim \mathcal{U}[1,64]$ for the mean function and $s_k \sim \mathcal{U}[1,8]$ for the geometry mean function. The normalized \gls{MSE} is averaged over $5 \times 10^4$ Monte Carlo trials.} 
  \label{fig:Sim_method}
   
\end{figure*}
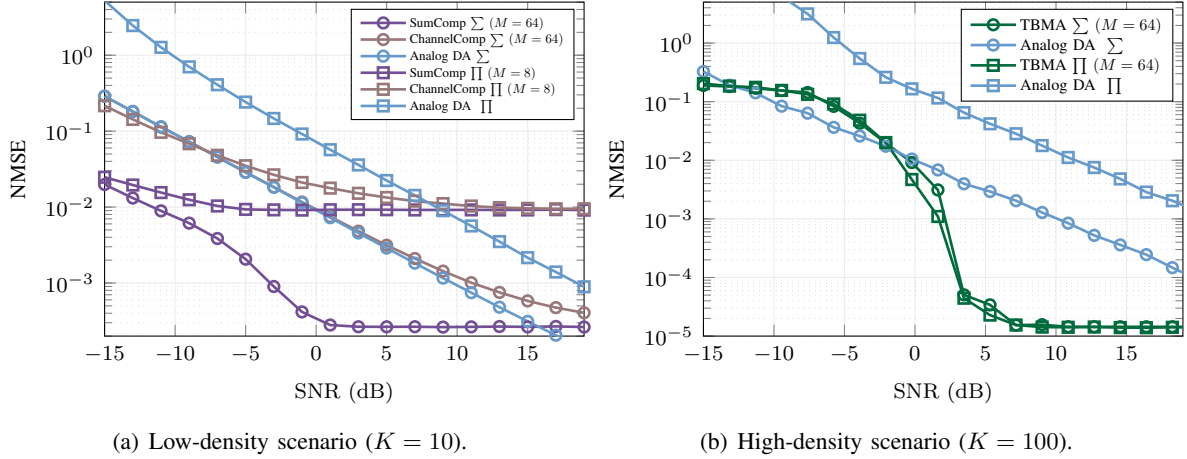

\subsection{Summary}
Table \ref{table:modulations} summarizes the modulations that have been used so far in the literature of \gls{AirComp}.
We highlight that analog angular modulations are generally not suitable for \gls{AirComp}, while existing digital angular modulations can be used with \gls{TBMA}. Furthermore, the development of new modulations for \gls{AirComp} is a promising field of research. Use case $4$ illustrates the integration of modulations for the application of \gls{AirComp} in \gls{FEEL}.

Fig. \ref{fig:Sim_method} compares the performance of the presented modulations for two different functions, namely, the arithmetic and geometric means. Fig~\ref{fig:Sim_low} shows the performance of the hybrid modulations ChannelComp~\cite{razavikia23_channelcomp} and SumComp~\cite{razavikia2023sumcomp} versus analog \gls{SSB} in terms of normalized MSE in a low-density scenario of $K=10$ users. All these modulations follow a DA scheme. The superiority of SumComp over ChannelComp is attributed to the fact that the Euclidean distance constraints in ChannelComp \eqref{eq:nonconvex_modulation}  is not the optimal choice, as it neglects the noise distribution in the channel. ChannelComp results in PAM modulations as the optimal solution for computing the sum function rather than QAM modulation. Although ChannelComp posits that PAM is superior to QAM, this assumption generally holds for uniform distributions but not for the AWGN channel.
Fig.~\ref{fig:Sim_high} compares \gls{TBMA} to analog \gls{SSB} in a high-density scenario of $K=100$ nodes.
Since the presented hybrid modulations change the constellation depending on the number of nodes $K$, and the optimization problem becomes harder to solve for an increasing $K$, it best suits scenarios with a low density of users. Conversely, the configuration of \gls{TBMA} is independent of $K$ and highly benefits from similar measurements, which suits scenarios with a high density of users.


\begin{table}
	\centering
\caption{Summary of modulations for \gls{AirComp}.}	
\resizebox{\textwidth*3/4-0.1in}{!}{
	\begin{tabular}{l|c|c|c|c}
		Modulation  & Category & Multiple Access & References  \\
		\hline\hline \begin{tabular}{@{}l@{}}\gls{SSB}\end{tabular} & Amplitude & \gls{DA} & None  \\
		\hline \begin{tabular}{@{}l@{}}Analog \gls{PAM}\end{tabular} & Amplitude & \gls{DA} & \cite{zhu20_baa, goldenbaum13_robust, cao20_powercontrol, gold1}  \\		
		\hline\hline \begin{tabular}{@{}l@{}}Digital \gls{PAM} \end{tabular}& Amplitude & \gls{DA} & \cite{wu16_stac}  \\	
		\hline \begin{tabular}{@{}l@{}}\gls{FSK} \end{tabular}& Angular & \gls{TBMA} & \cite{xie2023joint, martinez23_tbma, sahin22_numerals}\\		
		\hline \begin{tabular}{@{}l@{}}Digital \gls{PPM} \end{tabular}& Angular & \gls{TBMA} & \cite{sahin23_noncoherent}\\		
		\hline \begin{tabular}{@{}l@{}}\gls{Log-FSK}\end{tabular} & Angular & \gls{DA} & \cite{martinez2024log} \\		
		\hline \begin{tabular}{@{}l@{}}Hybrid amplitude-\\angular\end{tabular} & Amplitude-angular & \gls{DA} & \cite{razavikia23_channelcomp, guirado23whype}\\
		\hline
	\end{tabular}
}
\label{table:modulations}
\end{table}

\subsection*{Use case $4$: Modulations for \gls{FEEL}}

\begin{figure}
\centering
{\definecolor{copperrose}{RGB}{74,144, 226}
\tikzset{every picture/.style={line width=0.75pt}} 

\begin{tikzpicture}[x=0.75pt,y=0.75pt,yscale=-1]
\draw (55,40) node  {\footnotesize Tx$1$ (Upvote):};
\draw (55,70) node  {\footnotesize Tx$2$ (Upvote):};
\draw (55,100) node  {\footnotesize Tx$3$ (Downvote):};

\draw[-latex]    (110,40) -- (210,40) ;
\draw[-latex]    (110,70) -- (210,70) ;
\draw[-latex]    (110,100) -- (210,100) ;
\draw[-latex]    (250,45) -- (290,65) ;
\draw  [dash pattern={on 4.5pt off 4.5pt}]  (240,20) -- (240,150) ;
\draw[-latex]    (250,75) -- (290,75) ;
\draw[-latex]    (250,105) -- (290,85) ;
\draw[dash pattern={on 0.84pt off 2.51pt}] (265,75) ellipse (5 and 30);
\draw[-latex]    (310,90) -- (410,90) ;
\draw    (315,25) -- (315,95) ;
\draw[-latex] [color=copperrose ]   (140,40) -- (140,20) ;
\draw   [draw opacity=0][fill=copperrose ] (140,40) circle (3);
\draw   [draw opacity=0][fill=copperrose ] (180,40) circle (3);

\draw[-latex] [color=copperrose ]   (140,70) -- (140,50) ;
\draw   [draw opacity=0][fill=copperrose ] (140,70) circle (3);
\draw   [draw opacity=0][fill=copperrose ] (180,70) circle (3);

\draw[-latex] [color=copperrose ]   (180,100) -- (180,80);
\draw   [draw opacity=0][fill=copperrose ] (140,100) circle (3);
\draw   [draw opacity=0][fill=copperrose ] (180,100) circle (3);
\draw (140,110) node   {\footnotesize $\nearrow$};
\draw (180,110) node   {\footnotesize $\searrow$};
\draw[-latex] [color=copperrose ]   (330,90) -- (330,50);
\draw   [draw opacity=0][fill=copperrose ] (330,90) circle (3);
\draw[-latex] [color=copperrose  ,draw opacity=1 ]   (370,90) -- (370,75) ;
\draw   [draw opacity=0][fill=copperrose ] (370,90) circle (3);
\draw (380,83) node   {\footnotesize $e_{-}$ };

\draw  [color=black!40, dash pattern={on 4.5pt off 4.5pt}]  (315,75) -- (380,75) ;
\draw  [color=black!20, dash pattern={on 4.5pt off 4.5pt}]  (315,50) -- (380,50) ;
\draw (340,58) node   {\footnotesize $e_{+}$ };
\draw (335,40) node   {\footnotesize Energy};
\draw  [dash pattern={on 4.5pt off 4.5pt}]  (20,120) -- (520,120) ;
\draw  [dash pattern={on 4.5pt off 4.5pt}]  (415,20) -- (415,150) ;

\draw (300,75) node  {\footnotesize Rx:};
\draw (222,40) node {\footnotesize bins};
\draw (222,70) node {\footnotesize bins};
\draw (222,100) node {\footnotesize bins};
\draw (130,130) node   {\footnotesize Step~$1$: Active bins based on the classes };
\draw (132,145) node   {\footnotesize(i.e., Up $\nearrow$ or Down $\searrow$) };
\draw (335,130) node {\footnotesize  Step $2$: Non-coherent};
\draw (350,145) node {\footnotesize   aggregation};

\draw (460,130) node  {\footnotesize Step $3$: Compare };
\draw (485,145) node  {\footnotesize the energy};

\draw (460,70) node  {$\underbrace{e_{+} > e_{-}}_{\substack{\text{Majority voting}\\ \text{decision: Up } \nearrow}}$};

\end{tikzpicture}}
\caption{An example of computing a quantized majority voting function with non-coherent \gls{AirComp}. Bins can be subcarrier indices (i.e., FSK), chirp indices (i.e., CSK), or indices in time (i.e., PPM).}
\label{fig:noncoherent_FEEL}
\end{figure}
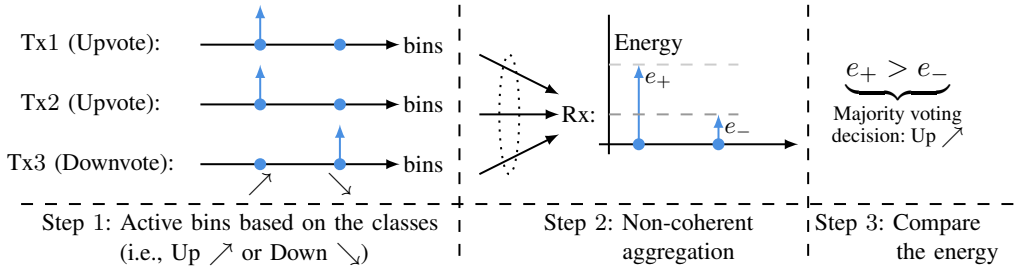

\gls{FEEL} is the leading application for \gls{AirComp} in the literature. See Section \ref{sec:applications} for a brief description of the operation of \gls{FEEL}. 
When the objective is to average the model parameters, where $f$ is the arithmetic mean, amplitude modulations are a viable approach~\cite{zhu20_baa}.
Alternatively, $M$-ary \gls{FSK} can be used to implement \gls{TBMA}~\cite{martinez23_feel}.
Furthermore, \gls{FEEL} can be realized by averaging gradients instead of model parameters.
In  \cite{zhu21_obda}, each gradient is quantized to a single bit indicating the sign of the gradient, and encoded into a \gls{BPSK} waveform. The receiver recovers the sign rather than the magnitude of the waveform, making $f$ a majority voting function, offering increased noise robustness and hardware efficiency. However, this approach still corresponds to an amplitude aggregation and requires \gls{CSI} to compensate for the channel effects.
To address these challenges, \cite{sahin23_noncoherent}, proposes  encoding the gradient sign with binary \gls{FSK} and implementing it using TBMA \gls{TBMA}. Computing the majority voting over binary \gls{TBMA} is achieved with an energy detector that compares both received signals (see Fig. \ref{fig:noncoherent_FEEL}). This approach enables a \gls{CSI}-free \gls{AirComp} scheme for \gls{FEEL}, enhancing its practicality and robustness.

\section{Physical layer frame}
\label{sec:physical_frame}

In its basic conception, the \gls{PHY} frame handles pulse shaping, carrier modulation and transmitted power control, so that the transmitted waveform $x_k^{\text{pb}}(t)$ is produced as\footnote{We are not considering in this model the case of event-based transmission, where $T$ is a random variable. $M$-ary continuous phase \gls{FSK} signals with $M>2$ are not encompassed by this model either.}:
\begin{align}
    x_k^{\text{pb}}(t) = \Re\{\sum_{n=1}^{\infty} A_c a_k[n] p(t-nT)\constante^{\constantj2\pi\fcarrierED[\indexED]\timeSymbol}\},
    \label{eq:passband}
\end{align}
where $A_c$ is the amplitude transmitted by the power control, $p(t)$ is the pulse shape of duration $T$ and unit energy, and the carrier frequency $f_{tx,k}$ is a integer multiple of $1/T$. However, the \gls{PHY} frame also constructs sophisticated modulation schemes for multiplexing or diversity. That is, the \gls{PHY} frame adapts the baseband symbols $a_k$ for passband transmission and makes them robust against impairments (e.g., interferences, multipath, noise,...). To achieve these, the \gls{PHY} frame uses extra radio resources to provide a processing gain at the receiver side. For instance, when $Q$ readouts in one terminal are to be multiplexed, a serial to parallel conversion takes place in the \gls{PHY} frame so that each one maps to a different carrier (i.e., Orthogonal Frequency Division Multiplex - OFDM) or, antenna (i.e., Multiple Input Multiple Output - MIMO) or code (i.e., Code Division Multiple Access - CDMA). Although under ideal transmission conditions, the \gls{PHY} frame and de-frame (i.e., at the receiver) processes must be transparent to the function computation, these schemes are revisited in next section due to their importance in \gls{AirComp} when dealing with the channel impairments. Note that in \gls{AirComp} the time synchronization and  \gls{CSI} requirements at each of the transmitters are more critical than in a conventional \gls{MAC} scheme, which does not aim to compute any function. We demonstrate the indispensability of OFDM schemes in easing these demands. As an alternative, a variation of CDMA, as outlined in \cite{goldenbaum13_robust}, is suggested to render the resultant scheme resilient against synchronization errors at both time and phase levels. In this approach, each readout $s_k$, whether analog or digital, modulates a random sequence of length $Q$. Besides, multi-antenna or MIMO signal processing and optimization must be differently approached in AirComp as next section will also point out. 
Note that multiplexing schemes are particularly useful in applications with large amounts of data (e.g., \gls{FEEL}) in order to reduce latency, and that whenever the frequency domain is used, the demodulator must implement \gls{TBMA}. 
Additionally, spread spectrum techniques are very useful in combating interference. In \cite{martinez23_feel}, the symbol $a_{k}$ is modulated in the frequency of a \gls{CSS} signal, which corresponds to the \gls{LoRa} modulation.  
Overall, the range of signal processing techniques to be devised in the \gls{PHY} frame for \gls{AirComp} is huge and constitutes quite a virgin research field.

\section{Challenges of multiple access channels for AirComp}
\label{sec:challenges}

In wireless communication networks, practical systems encounter challenges that significantly diverge from ideal models represented in \eqref{eq:mac_baseband_ideal}. These include hardware impairments, propagation distortions such as path loss, shadowing, multipath fading, thermal noise, and network interference. Such issues are aggravated by synchronization errors among the wireless units. Additionally, incorporating physical parameters and practical limitations into modeling processes is crucial to represent realistic wireless channels accurately.  In what follows,  we give a general system model for the fading \gls{MAC} with a detailed exploration of each term provided in subsequent subsections.  

In particular, let $x_k(t)\in\complexNumbers$ be the baseband signal for the $\indexED$th transmitter. Hence, the passband signal of the $\indexED$th transmitter that is formulated in \eqref{eq:passband} can also be expressed as
\begin{align}
	x_k^{\text{pb}}(t) = \Re\{x_k(t)\constante^{\constantj2\pi\fcarrierED[\indexED]\timeSymbol}\}.
    \label{eq:passbandtx}
\end{align} 
Assume that the impulse response of the multi-path channel in the passband is given by
\begin{align}
	h_k(t)=\sum_{\pathIndex=1}^{\numberOfPaths} h_{k,w}^c\delta(t-\tau_{k,w}),
\end{align}
where $\numberOfPaths$ is the number of paths, $h_{k,w}^c \in\realNumbers$ and $\tau_{k,w}\in\realNumbers$ are the $\pathIndex$th path gain and path delay for the $\indexED$th transmitter, respectively. The received passband signal $r^{\text{pb}}(t)$ can then be expressed as
\begin{align}
	r^{\text{pb}}(t)= \sum_{k=1}^{K} h_k(t)\circledast x_k^{\text{pb}}(t)+w(t).
		\label{eq:passbandrx}
\end{align}
The following subsections will explore the signal processing techniques that take place at the physical layer framing and at reception under different scenarios of fading effects and \gls{CSI} availability.


\subsection{Full \gls{CSI} availability}
\label{subsec:full CSI}
The following techniques specifically tailored for linear modulations. Without loss of generality, we refer this section to digital amplitude modulations, where each transmitter modulates $\hat\varphi_k(s_k)$, and \gls{DA} is the multiple access scheme. These modulated signals are then encapsulated within a physical layer frame to facilitate proper transmission and reception within the wireless channel.
Afterward, the received signal $r^{\text{pb}}(t)$ is translated to baseband, $r(t)$, sampled at symbol rate $T$ to produce $r[n]$, and further processed by the system $\mathcal{D}$, resulting in the output signals $y_{l}$, i.e., $y_l = \mathcal{D}\{r[n]\}$. 
Throughout this section, we assume that $f$ corresponds to the sum or arithmetic mean for $L=1$. To simplify exposition, the channels are assumed to have slow fading such that their states remain constant without the communication window. The extension to time-varying channels can be found in \cite{cao20_powercontrol}. 

\subsubsection{Power control}
\label{subsubsec:PowerControl}
In light of the previous assumptions, we can rewrite \eqref{eq:mac_baseband} as
\begin{equation}
    y = \sum_{k=1}^K h_kb_ks_k + w,
    \label{eq:model_precoder}
\end{equation}
where $b_k$ is the precoder. Note that \eqref{eq:model_precoder} assumes a linear model for the pre-processing function, namely, $\varphi_k(s_k)=b_ks_k$.
Channel-inversion is typically applied to achieve the magnitude alignment required for \gls{AirComp} \cite{zhu21_obda,zhu20_baa}. However, \textit{is channel inversion the optimal strategy for AirComp?} Considering an arbitrary symbol slot, this strategy is to set the transmit beamformer to 
\begin{align}
    b_k = \begin{cases}
        \rho_0 \bar h_{k}/|h_{k}|, \quad   |h_{k}|^2 \geq g_{\rm th}\\
        0, \quad\qquad \qquad |h_{k}|^2 < g_{\rm th}
    \end{cases},
    \label{eq:power_control_threshold}
\end{align}
where $\bar{h}_k$ is the complex conjugate of the channel response, $\rho_0$ is a scaling factor set to satisfy the transmit power, and $g_{\rm th}$ is a channel-truncation threshold to cope with deep fading. A quick answer to the initial question is negative in practice, where sensors are typically power-constrained. To be more elaborate, the optimal strategy exhibits a threshold-based structure such that a transmitter applies channel-inversion if its quality indicator,  which accounts for both its power budget and channel power gain, exceeds the threshold; otherwise, full power transmission should be performed \cite{cao20_powercontrol}. The detailed mathematical analysis and the resultant optimal strategy are presented in the sequel. 

If we set $\rho_0=\sqrt{p_k}$, where $p_k\geq0$ denotes the transmit power at device $k$, and perform channel inversion for \textit{all} users (i.e., the first case in \eqref{eq:power_control_threshold}), \eqref{eq:model_precoder} turns into 
\begin{align}
y=\sum_{k=1}^K\sqrt{p_{k}}|h_{k}| s_{k}+ w. \label{time_Yv}
\end{align}
Upon receiving the signal $y$, the receiver applies the denoising factor $\eta$, to recover the average message of interest as
$\hat f = y/(K\sqrt{\eta})$. Then, the MSE can be calculated as
\begin{equation}\label{MSE_ave}
 {\rm MSE}(f) \!=\!\frac{1}{K^2}~\mathbb{E}\!\left[\!\left(\!\frac{y}{\sqrt{\eta}}\!-\! \!\sum_{k=1}^K \!s_{k}\!\!\right)^2\!\, \right].
\end{equation}
Our objective is to minimize \eqref{MSE_ave} by jointly optimizing the power control $ p_{k}$ at devices and the denoising factor $\eta$ subject to the individual average transmit power constraints. This results in the following power control problem:
\begin{subequations}
\begin{align}
\nonumber
& \underset{\{ p_k\geq0\},\eta\geq0}{\text{minimize}} ~~~
\sum \limits_{k=1}^K\left(\frac{\sqrt{p_{k}}|h_{k}|}{\sqrt{\eta}}-1\right)^2+\frac{\sigma^2}{\eta} \\
& \hspace{-20pt}\qquad\text{subject to}~~~~
p_{k}\leq {P}_k, ~\forall k
\end{align}
\label{eq:power_control}
\end{subequations}
where ${P}_k$ is the average power budget. It is observed that the objective function of problem \eqref{eq:power_control} consists of two components: the signal misalignment error and the noise-induced error, respectively.
In general, adjusting  $\eta$ induces a fundamental tradeoff between minimizing the signal misalignment error and suppressing the noise-induced error.
It is evident that the optimal policy to problem \eqref{eq:power_control} for device $k$ is given by
\begin{align}
p_k^{*}=\min~\left({P}_k,\frac{\eta}{|h_k|^2}\right).\label{K_device_N1_Pk_opyt}
\end{align}
Next, we proceed to optimize over $\eta$. By substituting $p_k^{*}$ in \eqref{K_device_N1_Pk_opyt} into  problem \eqref{eq:power_control}, we have the optimization problem over $\eta$ as
\begin{align}
\underset{\eta\geq 0}{\text{minimize}}
~\sum_{k=1}^K \left(\min\left(\frac{\sqrt{{P}_k}|h_{k}|}{\sqrt{\eta}}-1,0\right)\right)^2+\frac{\sigma^2}{\eta},
\label{static_K_eta}
\end{align}
Even though a closed-form solution is difficult, its structure can be characterized by studying the properties of the objective function, e.g., its unimodality \cite{cao20_powercontrol}. To solve problem \eqref{static_K_eta}, we need to remove the minimum operation to simplify the derivation. To this end, we find it convenient to adopt a \emph{divide-and-conquer} approach that divides the feasible set of problem \eqref{static_K_eta}, namely $\{\eta\geq 0\}$, into $K+1$ intervals. Note that each of them is defined as
\begin{align}
	{\cal F}_k\!=\!\{\eta \mid {P}_k|h_k|^2 \!\leq \eta \leq \!{P}_{k+1}|h_{k+1}|^2\!\}, ~\forall k\in \{0\}\cup\cal K,
\end{align}
where we define $\mathcal{K}=\{1,\dots,K\}$, ${P}_{0}|h_{0}|^2\triangleq0$ and ${P}_{K+1}|h_{K+1}|^2 \rightarrow \infty$ for notational convenience.
Then, it is easy to establish the equivalence between the following two sets:
\begin{align}\label{static_eta_spec_domian}
\{\eta\geq 0\}=\bigcup \limits_{k\in \{0\}\cup\cal K} {\cal F}_k.
\end{align}
Given \eqref{static_eta_spec_domian}, we note that solving problem \eqref{static_K_eta} is equivalent to solving the following $K+1$ subproblems and comparing their optimal values to obtain the minimum one:
\begin{align}
\underset{\eta\in{\cal F}_k}{\text{minimize}}~
F_k(\eta)\! \triangleq \!\sum_{i=1}^k\! \left(\!\frac{\sqrt{{P}_i}|h_{i}|}{\sqrt{\eta}}\!-\!1\!\!\right)^2\!+\!\frac{\sigma^2}{\eta}, ~\forall k\in\{0\}\cup\cal K,\!
\label{eq:P2}
\end{align}
with $F_k(\eta)$ denoting the objective function of the $k$th subproblem. Notice that we have $F_0(\eta) =  \sigma^2/\eta$  for $k =0$. In addition, we have
\begin{align}\label{static_F_Fk}
	F(\eta) = F_k(\eta), ~\forall \eta \in {\cal F}_k,~&\forall k\in\{0\}\cup\cal K.
\end{align}
With the optimal index $k^*$ that can be found by the search, the optimal precoder over static channels that solves problem \eqref{eq:P2} has a threshold-based structure, given by
\begin{align}
	p_k^*=
	\begin{cases}
		{P}_k,~&\forall k\in\{1,\cdots,k^* \},\\
		\frac{\eta^*}{|h_{k}|^2},~&\forall k\in\{k^*+1,\cdots,K \},
	\end{cases}\label{static_p_lemma}
	\end{align}
where the threshold is given as 	
\begin{align}	
	\eta^*=\tilde \eta_{k^*}=\left( \frac{\sigma^2+ \sum_{i=1}^{k^*} {P}_{i}|h_{i}|^2}{\sum_{i=1}^{k^*}\sqrt{{P}_{i}}|h_{i}|}\right)^2.\label{opt_eta_static}
\end{align}
Furthermore, it holds that ${P}_k|h_k|^2\leq\eta^* $ for devices $k\in\{1,\cdots,k^* \}$ and ${P}_k|h_k|^2\geq\eta^*$ for devices $k\in\{k^*+1,\cdots,K \}$. 
The optimal policy over devices has a threshold-based structure. The threshold is specified by the denoising factor $\eta^*$ and applied on the derived quality indicator $\bar{P}_{k}|h_{k}|^2$ accounting for both the channel power gain and power budget of device $ k\in\cal K$. It is shown that for each device $k\in \{k^*+1,..., K\}$ with its quality indicator exceeding the threshold, i.e., ${P}_k|h_k|^2\geq\eta^*$, the \emph{channel-inversion} is applied with $p_k^* = \frac{\eta^*}{|h_{k}|^2}$; while for each device $k \in \{1, ..., k^*\}$ with ${P}_k|h_k|^2\leq\eta^*$, the \emph{full power transmission} is deployed with $p_k^*=P_k$. 

\vspace{5mm}
\subsubsection{MIMO AirComp}
In a communication system equipped with multiple antennas, spatial multiplexing enables the simultaneous transmission of parallel data streams without requiring additional bandwidth. In the context of AirComp, spatial multiplexing can be leveraged to enhance computational throughput or support vector-function computation, with the multiplexing gain linearly increasing with antenna sizes. The result is termed MIMO AirComp. Some spatial degrees of freedom (DoF) can also be allocated to minimizing the AirComp error by exploiting spatial diversity.

In a single-antenna system, the conventional technique for scalar-function AirComp is relatively simple as it mostly involves channel inversion at each transmitter (e.g., sensor). The techniques for MIMO AirComp is sophisticated. While channel inversion remains optimal as implemented as zero-forcing beamforming, the optimization of receive beamforming, called aggregation beamforming, is NP-hard, as shown in the sequel. A standard approach can be applied to solve the problem based on \emph{semidefinite programming} (SDP) to design an efficient iterative interior point algorithm. But little insight can be derived. Researchers have found a close-to-optimal solution revealing that the optimal aggregation beamforming is a weighted centroid of multiuser MIMO channels projected onto a Grassmann manifold (or simply called a Grassmannian). This will be the focus of our exposition in the remainder of this section.

To formulate the optimization problem for aggregation beamforming, let us introduce some additional notation. Let $\mathbf{s}_k\in\mathbb{C}^Q$ be a $Q$-length baseband signal transmitted by user $k$, and $\bB_k \in \mathbb{C}^{N_t\times Q}$ the transmit beamforming matrix at transmitter $k$. $N_t$ and $N_r$ are the number of transmit and receive antennas, respectively. The signal at the input of the receiver is $\mathbf{y}\in\mathbb{C}^{N_r}$ and the signal model in \eqref{eq:mac_baseband} turns into
\begin{align}\label{estimated_y}
\mathbf{y} = \sum_{k=1}^K \bH_k \bB_k \bs_k + \bn,
\end{align}
where $\bH_k  \in \mathbb{C}^{N_r\times N_t}$ represents the MIMO channel matrix for the link from the transmitter $k$ to the receiver, and $\bn$ is noise vector distributed as $\mathcal{CN}(0, \sigma_n^2)$ elements.

The ultimate goal of the \gls{AirComp} MIMO scheme is not to recover each individual $\mathbf{s}_k$, but the sum $\sum_{k=1}^K \mathbf{s}_k$. For this reason, the received signal is further processed with a receive beamforming matrix $\bA \in \mathbb{C}^{N_r\times Q}$: 
\begin{align}\label{estimated_s}
\hat \bs = \bA^\text{H} \by=
\bA^\text{H}\sum_{k=1}^K \bH_k \bB_k \bs_k + \bA^\text{H}\bn,
\end{align}
where $(\cdot)^\text{H}$ is the Hermitian of a matrix. 
We define the desired function as $\bff = \sum_k \bs_k $ and its noisy version $\hat{\bff} = \hat \sum_k \bs_k $. Then the  AirComp error is defined  by the following MSE:
\begin{equation}\label{def:MSE}
 \text{MSE}({\hat{\bf f}}) =\mathbb{E} \l [\tr\l(\sum_{k=1}^K \bs_k - \sum_{k=1}^K \hat{\bs}_k\r)\l(\sum_{k=1}^K\bs_k - \sum_{k=1}^K \hat{\bs}_k\r)^\text{H}\r],
\end{equation}
where $\tr(\cdot)$ is the trace operator. 
Substituting \eqref{estimated_s} into \eqref{def:MSE}, the MSE  can be explicitly written as a function of the transmit and receive beamformer as follows:
\begin{equation}\label{MSE_function}
 \!\! \text{MSE}(\bA, \{\bB_k\}) \!=\! \sum_{k=1}^K \tr\left( (\bA^\text{H}\bH_k \bB_k - \bI) (\bA^\text{H}\bH_k \bB_k - \bI)^\text{H} \right) \\ + \sigma_n^2 \tr\left(\bA^\text{H}\bA\right).
\end{equation}

Consider the joint optimization of the transmit and aggregation beamformers to minimize the \gls{MSE}. As transmitters are typically sensors with limited power, it is necessary to set a constraint on the average transmission power of each transmitter such that
$\|\bB_k\|^2 \leq P_0,~\forall k$. Following a common approach in the MIMO beamforming literature, the receive beamformer $\bA$ is constrained to be an orthonormal matrix. Moreover, under the MMSE (i.e., minimum AirComp error) criterion, the denoising factor (positive scalar) can be absorbed into  $\bA$ for regulating the tradeoff between noise reduction and transmission-power control. We can write $\bA = \sqrt{\eta} \bF$ with $\bF$ being a tall unitary matrix and thus $\bF^\text{H}\bF = \bI$. 
Then given the AirComp error in \eqref{MSE_function},  the aggregation  beamforming problem can be  formulated as:
\begin{subequations}
\begin{align}
\nonumber
& \underset{\eta, \bA, \{\bB_k\}}{\text{minimize}} ~~~~
\text{MSE}(\bA, \{\bB_k\}) \\
& \hspace{-20pt}\qquad\text{subject to}~~~
\|\bB_k\|^2 \leq P_0, \;\forall k\\
& ~~~~~~~~~~~~~~~
\bA^\text{H} \bA = \eta \bI
\end{align}
\label{eq:MIMO_beamformer}
\end{subequations}
Problem \eqref{eq:MIMO_beamformer} is difficult to solve due to its non-convexity. The lack of convexity arises from the coupling between transmit and receive beamformers in the objective function, and the orthogonality constraint on the receive beamformer. 
It can be proved that zero-forcing transmit beamforming conditioned on an aggregation  beamformer, as given below, is optimal as follows \cite{zhu19_mimo}
\begin{align}\label{ZF}
\bB_k^* = (\bA^\text{H} \bH_k)^\text{H} (\bA^\text{H} \bH_k \bH_k^\text{H} \bA)^{-1}.
\end{align}
As a result, Problem \eqref{eq:MIMO_beamformer} is  transformed into the equivalent problem of minimizing the denoising factor of the receive beamformer:
\begin{subequations}
\begin{align}
\nonumber
& \underset{\eta, \bF}{\text{minimize}} ~~~~
\eta \\
& \hspace{-20pt}\qquad\text{subject to}~~~
\frac{1}{\eta} \text{tr} \left((\bF^\text{H} \bH_k \bH_k^\text{H} \bF)^{-1}\right) \leq P_0,~ \forall k\\
& ~~~~~~~~~~~~~~~
\bF^\text{H} \bF = \bI
\end{align}
\label{eq:MIMO_denoising}
\end{subequations}
where $\bF$ is defined earlier as the normalized receive beamformer. To develop a tractable approximation of the problem, we can modify the power constraints with an approximate form. This requires  \emph{singular value decomposition}  (SVD) of $\bH_k$ of each MIMO channel: $\bH_k = \bU_k \Sigma_k \bV_k^\text{H}$. As a result, Problem \eqref{eq:MIMO_denoising} can be approximated by the following problem with tightened power constraints and   invoking differential geometry theory to reveal its equivalence to the following form that contains subspace distances between the  aggregation  beamformer and individual MIMO channels (see details in \cite{zhu19_mimo})
\begin{subequations}
\begin{align}
\nonumber
& \underset{\bF}{\text{maximize}} ~~~~
\sum_{k=1}^K \lambda_{\min}(\Sigma_k^2) \text{tr}(\bU_k^\text{H} \bF \bF^\text{H} \bU_k) \\
& \hspace{-20pt}\qquad\text{subject to}~~~
\bF^\text{H} \bF = \bI\\
& ~~~~~~~~~~~~~~~
d_{\sf PF}^2 (\bU_k, \bF) = N_t - \text{tr}(\bU_k^\text{H} \bF \bF^\text{H} \bU_k)\label{eq:ineq}
\end{align}
\label{eq:MIMO_denoising_eigen}
\end{subequations}
Problem \eqref{eq:MIMO_denoising_eigen} remains non-convex due to maximizing a convex objective function and the orthogonality constraints on the variable $\bF$. Nevertheless, by intelligently constructing an equivalent unconstrained problem,  we can derive a closed-form solution by analyzing the stationary points of the unconstrained problem. For ease of exposition, define a matrix   $\bG \in \mathbb{C}^{N_r \times N_r}$ as follows: 
\begin{align}\label{CSI_function}
\bG = \sum_{k=1}^K \lambda_{\min}(\Sigma_k^2) \bU_k \bU_k^\text{H}.
\end{align}
Let $\bG = \bV_{G} \Sigma_G \bV_{G}^\text{H}$ be the SVD of $\bG$ given in \eqref{CSI_function}. The solution is given by  the first $L$ principal eigenvectors of $\bG$, namely
\begin{align}\label{Eq:RxBeam}
\bF^* = [\bV_G]_{:,1:Q}.
\end{align}
It follows that the optimal aggregation beamforming matrix  is given as 
$\bA^* = \sqrt{\eta^*} \bF^*$,
while the corresponding denoising factor is given as 
\begin{equation}
\eta^* = \underset{k}{\text{maximize}}~ \frac{1}{P_0} \text{tr} \left(((\bF^*)^\text{H} \bH_k \bH_k^\text{H} \bF^*)^{-1}\right). 
\end{equation}

Let us interpret the geometry structure of the above optimal aggregation beamformer. Each unitary matrix $\bU_k$ in \eqref{CSI_function} represents the spatial orientation of MIMO channel $k$. $\bU_k$ appears as a single point on a Grassmann manifold, which can be interpreted as the space of subspaces. As a result, $\bG$ is a weighted sum of MIMO channels projected onto the Grassmann manifold with weights related to the channels’ eigenvalues (or equivalently their norms). Being a derivative of $\bG$, the aggregation beamformer $\mathbf{A}^*$ allows the same geometric interpretation.

\subsection{Partial \gls{CSI}: Synchronization impairments}
\label{subsec:SyncImpairments}

In practice, transmitted signals are exposed to hardware imperfections in addition to the distortions in the propagation medium. While some of these impairments,  such as power amplifier non-linearity, can be single-sided, i.e., occurring only at transmitters, the distortions like timing errors due to the imperfect clock feeding the digital circuit, residual synchronization errors in the baseband, carrier frequency offset, phase offset, and phase noise in RF front-end, can occur at both transmitter and receiver. Most of these impairments can be addressed well in communication scenarios as the distortion can be estimated and corrected, particularly at the receiver. However, the same methods used for communications may not be directly applied to computation as the receiver observes the superposed distorted transmitted signals in AirComp, resulting in a new plethora of research topics and new AirComp schemes for reliable computations. Next, we assess the impact of synchronization impairments in AirComp. Unlike in the previous subsection \ref{subsec:full CSI}, the signal processing techniques applied in this section are not constrained to amplitude modulations. Furthermore, we assume $L\geq1$ and a single antenna per transmitter and receiver.

\input{images/Fig_synch}
Let $\{\fcarrierED[\indexED],\phaseED[\indexED]\}$ and $\{\fcarrierES,\phaseES\}$ be the carrier frequency and the phase of the \gls{LO} at the $\indexED$th transmitter and the receiver, respectively, as shown in \figurename~\ref{fig:syncDiagram}, and further completing the formulation of the transmitted passband signals in \eqref{eq:passbandtx} as
\begin{align}
	x_k^{\text{pb}}(t) = \Re\{x_k(t)\constante^{\constantj2\pi\fcarrierED[\indexED]\timeSymbol+j\phaseED[\indexED]}\}.
\end{align} 
The received passband signal for the $k$th transmitter can then be expressed as
\begin{align}
	r_k^{\text{pb}}(t)=  h_k(t)\circledast x_k^{\text{pb}}(t)=\Re\left\{\constante^{\constantj2\pi\fcarrierES\timeSymbol+\constantj\phaseES}\underbrace{\sum_{\pathIndex=1}^{\numberOfPaths}
		\gainComplex[\indexED,\pathIndex]\constante^{\constantj\po[\indexED]}
		x_k({\timeSymbol-\delay[\indexED,\pathIndex]})\constante^{\constantj2\pi\cfo[\indexED](\timeSymbol-\delay[\indexED,\pathIndex])}}_{\triangleq r_k(t)}\right\},
		\label{eq:genericSync}
\end{align}
where $\cfo[\indexED]\triangleq\fcarrierED[\indexED]-\fcarrierES$, and $\po[\indexED]\triangleq\phaseED[\indexED]-\phaseES$ and the \gls{CFO} are the \gls{PO} between the $\indexED$th transmitter and the receiver, respectively, and $r_k(t)$ is the received complex baseband signal for 
$h_{k,p}(t)\triangleq h_{k,p}^c\constante^{-\constantj2\pi\fcarrierES\delay[\indexED,\pathIndex]}$. 

Now, suppose that the following conditions hold for all transmitters: 
\begin{itemize}
	\item Condition 1: $\po[\indexED]$ is not a function of time. Condition 1 implies that the transmitters and receiver do not power down their LOs at any time. It also requires very accurate crystals; otherwise, $\po[\indexED]$ becomes random.
	\item Condition 2: The {\em baseband} signal $x_k(t)$ is offset by  $\to[\indexED]$  relative to the receiver’s synchronization point $\syncPoint$ and offsets are not jittery. Condition 2 does not consider potential jitters in the passband signal, which can lead to an additional phase rotation that is very sensitive to the value of carrier frequency. 
	\item Condition 3: Frequency synchronization is ideal, i.e., $\cfo[\indexED]=0$.	If Condition~3 is violated (e.g., residual CFO after correction), phase error accumulation over time occurs, as seen from \eqref{eq:genericSync}. Hence, coherent AirComp can be very sensitive to channel aging.
\end{itemize}

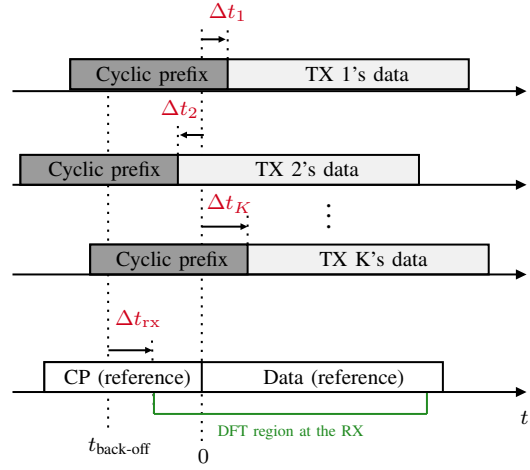
\begin{figure}[t]
    \centering

\scalebox{1}{

\tikzset{every picture/.style={line width=0.75pt}} 

\begin{tikzpicture}[x=0.75pt,y=0.75pt,yscale=-1,xscale=1]

\draw  [dash pattern={on 0.84pt off 2.51pt}]  (121.5,22) -- (121.5,229) ;

\draw  [dash pattern={on 0.84pt off 2.51pt}]  (97.5,215) -- (96.64,183.05) -- (96.5,178) ;
\draw  [dash pattern={on 0.84pt off 2.51pt}]  (74.5,51) -- (74.5,224) ;
\draw  [fill={rgb, 255:red, 241; green, 241; blue, 241 }  ,fill opacity=1 ] (55.5,37) -- (255.5,37) -- (255.5,52) -- (55.5,52) -- cycle ;
\draw  [fill={rgb, 255:red, 155; green, 155; blue, 155 }  ,fill opacity=1 ] (55.5,37) -- (134.5,37) -- (134.5,52) -- (55.5,52) -- cycle ;

\draw  [fill={rgb, 255:red, 255; green, 255; blue, 255 }  ,fill opacity=1 ] (42.5,189) -- (242.5,189) -- (242.5,204) -- (42.5,204) -- cycle ;
\draw  [fill={rgb, 255:red, 255; green, 255; blue, 255 }  ,fill opacity=1 ] (42.5,189) -- (121.5,189) -- (121.5,204) -- (42.5,204) -- cycle ;

\draw  [fill={rgb, 255:red, 241; green, 241; blue, 241 }  ,fill opacity=1 ] (65.5,130) -- (265.5,130) -- (265.5,145) -- (65.5,145) -- cycle ;
\draw  [fill={rgb, 255:red, 155; green, 155; blue, 155 }  ,fill opacity=1 ] (65.5,130) -- (144.5,130) -- (144.5,145) -- (65.5,145) -- cycle ;

\draw  [fill={rgb, 255:red, 241; green, 241; blue, 241 }  ,fill opacity=1 ] (30.5,85) -- (230.5,85) -- (230.5,100) -- (30.5,100) -- cycle ;
\draw  [fill={rgb, 255:red, 155; green, 155; blue, 155 }  ,fill opacity=1 ] (30.5,85) -- (109.5,85) -- (109.5,100) -- (30.5,100) -- cycle ;

\draw    (121.5,27) -- (130.5,27) ;
\draw [shift={(133.5,27)}, rotate = 180] [fill={rgb, 255:red, 0; green, 0; blue, 0 }  ][line width=0.08]  [draw opacity=0] (3.57,-1.72) -- (0,0) -- (3.57,1.72) -- cycle    ;
\draw  [dash pattern={on 0.84pt off 2.51pt}]  (134.5,37) -- (134.5,22) ;
\draw  [dash pattern={on 0.84pt off 2.51pt}]  (109.5,85) -- (109.5,70) ;
\draw    (122.5,75) -- (113.5,75) ;
\draw [shift={(110.5,75)}, rotate = 360] [fill={rgb, 255:red, 0; green, 0; blue, 0 }  ][line width=0.08]  [draw opacity=0] (3.57,-1.72) -- (0,0) -- (3.57,1.72) -- cycle    ;
\draw  [dash pattern={on 0.84pt off 2.51pt}]  (144.5,130) -- (144.5,115) ;
\draw    (121.5,121) -- (141.5,121) ;
\draw [shift={(144.5,121)}, rotate = 180] [fill={rgb, 255:red, 0; green, 0; blue, 0 }  ][line width=0.08]  [draw opacity=0] (3.57,-1.72) -- (0,0) -- (3.57,1.72) -- cycle    ;
\draw    (74.64,183.05) -- (93.64,183.05) ;
\draw [shift={(96.64,183.05)}, rotate = 180] [fill={rgb, 255:red, 0; green, 0; blue, 0 }  ][line width=0.08]  [draw opacity=0] (3.57,-1.72) -- (0,0) -- (3.57,1.72) -- cycle    ;
\draw [color={rgb, 255:red, 34; green, 139; blue, 34 }  ,draw opacity=1 ]   (97.5,215) -- (234.5,215) ;
\draw [color={rgb, 255:red, 34; green, 139; blue, 34 }  ,draw opacity=1 ]   (234.5,204) -- (234.5,215) ;
\draw [color={rgb, 255:red, 34; green, 139; blue, 34 }  ,draw opacity=1 ]   (97.5,204) -- (97.5,215) ;
\draw    (26.5,100) -- (281.5,100) ;
\draw [shift={(284.5,100)}, rotate = 180] [fill={rgb, 255:red, 0; green, 0; blue, 0 }  ][line width=0.08]  [draw opacity=0] (5.36,-2.57) -- (0,0) -- (5.36,2.57) -- cycle    ;
\draw    (26.5,53) -- (281.5,53) ;
\draw [shift={(284.5,53)}, rotate = 180] [fill={rgb, 255:red, 0; green, 0; blue, 0 }  ][line width=0.08]  [draw opacity=0] (5.36,-2.57) -- (0,0) -- (5.36,2.57) -- cycle    ;
\draw    (26.5,145) -- (281.5,145) ;
\draw [shift={(284.5,145)}, rotate = 180] [fill={rgb, 255:red, 0; green, 0; blue, 0 }  ][line width=0.08]  [draw opacity=0] (5.36,-2.57) -- (0,0) -- (5.36,2.57) -- cycle    ;
\draw    (26.5,204) -- (281.5,204) ;
\draw [shift={(284.5,204)}, rotate = 180] [fill={rgb, 255:red, 0; green, 0; blue, 0 }  ][line width=0.08]  [draw opacity=0] (5.36,-2.57) -- (0,0) -- (5.36,2.57) -- cycle    ;

\draw (67,39) node [anchor=north west][inner sep=0.75pt]  [font=\scriptsize] [align=left] {Cyclic prefix};
\draw (172,39) node [anchor=north west][inner sep=0.75pt]  [font=\scriptsize] [align=left] {TX 1's data};
\draw (147,87) node [anchor=north west][inner sep=0.75pt]  [font=\scriptsize] [align=left] {TX 2's data};
\draw (43,87) node [anchor=north west][inner sep=0.75pt]  [font=\scriptsize] [align=left] {Cyclic prefix};
\draw (51,191) node [anchor=north west][inner sep=0.75pt]  [font=\scriptsize] [align=left] {CP (reference)};
\draw (151,191) node [anchor=north west][inner sep=0.75pt]  [font=\scriptsize] [align=left] {Data (reference)};
\draw (118,231.4) node [anchor=north west][inner sep=0.75pt]  [font=\scriptsize]  {$0$};
\draw (63,225.4) node [anchor=north west][inner sep=0.75pt]  [font=\scriptsize]  {$t_{\text{back-off}}$};
\draw (280,211.99) node [anchor=north west][inner sep=0.75pt]  [font=\scriptsize]  {$t$};
\draw (123.5,7.4) node [anchor=north west][inner sep=0.75pt]  [font=\scriptsize,color={rgb, 255:red, 208; green, 2; blue, 27 }  ,opacity=1 ]  {$\Delta t_{1}$};
\draw (99.5,56.4) node [anchor=north west][inner sep=0.75pt]  [font=\scriptsize,color={rgb, 255:red, 208; green, 2; blue, 27 }  ,opacity=1 ]  {$\Delta t_{2}$};
\draw (122.5,103.4) node [anchor=north west][inner sep=0.75pt]  [font=\scriptsize,color={rgb, 255:red, 208; green, 2; blue, 27 }  ,opacity=1 ]  {$\Delta t_{K}$};
\draw (182,100) node [anchor=north west][inner sep=0.75pt]    {$\vdots $};
\draw (77,132) node [anchor=north west][inner sep=0.75pt]  [font=\scriptsize] [align=left] {Cyclic prefix};
\draw (179,132) node [anchor=north west][inner sep=0.75pt]  [font=\scriptsize] [align=left] {TX K's data};
\draw (128,219) node [anchor=north west][inner sep=0.75pt]  [font=\tiny,color={rgb, 255:red, 34; green, 139; blue, 34 }  ,opacity=1 ] [align=left] {DFT region at the RX};
\draw (76.5,161.4) node [anchor=north west][inner sep=0.75pt]  [font=\scriptsize,color={rgb, 255:red, 208; green, 2; blue, 27 }  ,opacity=1 ]  {$\Delta t_{\rm rx}$};

\end{tikzpicture}
}
    \caption{Time synchronization errors at the transmitter and receivers. The receiver backs off by $\syncPoint$~seconds for the DFT window to avoid samples from the subsequent OFDM symbol in practice while it can still be out-of-sync by $\Delta t_{\rm rx}$~second like the transmitters.}
    \label{fig:OFDM_timing}
\end{figure}
Considering the widespread adoption of OFDM in wireless networks, its robustness against multipath channels, and its compatibility with a wide range of waveforms via transform precoders (e.g., DFT-precoder for a single-carrier waveform),  suppose that symbols for \gls{AirComp} are transmitted with an \gls{OFDM} symbol of duration $T_{sym}$ as
\begin{align}
	x_k(t)=\begin{cases}
		\displaystyle\sum_{\indexTXsubcarrier=0}^{Q-1}a_{kq}\constante^{\constantj2\pi\indexTXsubcarrier\frac{\timeSymbol-\to[\indexED]}{\symbolDuration}},& -\cpDuration+\to[\indexED]\le\timeSymbol<\symbolDuration+\to[\indexED]\\
		0, &\text{otherwise}
	\end{cases}
\end{align}
Under Conditions 1-3, we can then express the superposed waveform for $\timeSymbol\in[\syncPoint,\symbolDuration-\syncPoint)$, i.e., the \gls{DFT} processing window at the receiver as shown in \figurename~\ref{fig:OFDM_timing}, as
\begin{align}
	r(t)
	=&\sum_{\indexRXsubcarrier=0}^{Q-1} 
	\underbrace{\sum_{\indexED=1}^{\numberOfEdgeDevices}\gainComplexCFR[\indexED,\indexRXsubcarrier]\constante^{\constantj\po[\indexED]}\constante^{\constantj2\pi\indexRXsubcarrier\frac{\syncPoint-\so-\to[\indexED]}{\symbolDuration}}
	a_{k,l}}_{\text{superposed symbols in the frequency domain}}\constante^{\constantj2\pi\indexRXsubcarrier\frac{\timeSymbol}{\symbolDuration}},
	\label{eq:finalsum}
\end{align}
for $\gainComplexCFR[\indexED,\indexRXsubcarrier]\triangleq\sum_{\pathIndex=1}^{\numberOfPaths}
h_{k,p}^c\constante^{\constantj2\pi\indexRXsubcarrier\frac{\delay[\indexED,\pathIndex]}{\symbolDuration}}$, i.e., the frequency response of the multipath channel. We can now infer the following results:
\begin{itemize}
    \item Under Conditions~1-3, the phase term in \eqref{eq:finalsum} due to the offsets increases with the subcarrier index~$\indexRXsubcarrier$ for $\syncPoint-\so-\to[\indexED]\neq0$. 
    \item Equation~\eqref{eq:finalsum} shed light on AirComp based on single-carrier waveform. The phase term in \eqref{eq:finalsum} due to the time misalignment is zero for the DC subcarrier, i.e., $\indexRXsubcarrier=0$, corresponding to a single carrier waveform with a rectangular pulse.
    \item The main issue for coherent \gls{AirComp} is not \gls{TO} or \gls{PO} themselves, but the jitters on \gls{TO} or \gls{PO} under Conditions~1-3. This is because a channel estimator estimates the composite response of the channel, i.e., $\gainComplexCFR[\indexED,\indexRXsubcarrier]\constante^{\constantj\po[\indexED]}\constante^{\constantj2\pi\indexRXsubcarrier\frac{\syncPoint-\so-\to[\indexED]}{\symbolDuration}}$, rather than the solely multi-path channel. Hence, it is crucial to keep the \glspl{PO} and \glspl{TO} as fixed as possible for coherent \gls{AirComp}. 
\end{itemize}

Conditions 1-3 introduce stringent design constraints for practical wireless networks. Hence, a fundamental challenge is how to achieve \gls{AirComp} without relying on the availability of phase synchronization in the network or the availability of instantaneous \gls{CSI} at the nodes, leading to various non-coherent approaches. 
One of the key ideas exploited in literature is to dedicate different resources representing a set of classes and use the energy accumulation on these resources to obtain an approximate histogram, i.e., \gls{TBMA}, discussed in Section~III.
For example, in a binary format, two classes may defined as 0 and 1, and a non-coherent AirComp can encode the binary information for each digit and the sum of digits can be exploited as discussed in Section VI.1.b.

Another key idea is to modulate the norm-square of the sequence with the magnitude of the parameter, proposed in \cite{goldenbaum13_robust}. This approach is different from the aforementioned TBMA-based non-coherent approaches in that it achieves continuous-valued aggregation while still exploiting the energy accumulation in the medium. In this approach, the output of the pre-processing function is processed further with a function $\affineEnc$ that results in a non-negative value. Subsequently, the {\em square root} of the resulting value is multiplied with  
a sequence of length $Q$ as $\sqrt{g(\parameter[\indexED])}\times[
		{\constante^{\constantj\theta_1}} ,\dots, {\constante^{\constantj\theta_Q}}]$, i.e., the norm-square of the sequence linearly changes with $g(\parameter[\indexED])$.  At the receiver, the energy of the received sequence is calculated to achieve the \gls{AirComp}, and the superposed symbol is processed with another affine function $\affineDec$ to reverse the impact of $\affineEnc$ on the superposed symbols. It is worthwhile noting that the affine function at the receiver is a function of the number of transmitters participating in the computation.
As the receiver calculates the energy of the received sequence with this technique, the scheme is not sensitive to phase synchronization errors. However, the price paid is the cross-product terms due to the loss of orthogonality of the sequences in an overcomplete space for AirComp with limited resources.  In  \cite{goldenbaum13_robust}, unimodular sequences with random phases are adopted to reduce the cross-product terms, but it is emphasized that the design of the sequence should harness interference as a common goal instead of eliminating the interference as in a traditional code division multiple access systems. Currently, the optimal sequence set for Goldenbaum's approach is an open question.

\begin{figure}
\centering
\begin{minipage}{.45\textwidth}
  \centering
     \begin{tikzpicture} 
    \begin{axis}[
        xlabel={Maximum phase deviation [degree] },
        ylabel={MSE},
        label style={font=\tiny},
        legend cell align={left},
        tick label style={font=\tiny} , 
        width=\textwidth,
        height=6cm,
        xmin=0, xmax=90,
        ymin=1e-2, ymax=3,
        xtick={0, 10, 20, 30, 40,50,60,70,80,90},
        ymode = log,
       legend style={nodes={scale=0.55, transform shape}, at={(0.98,0.3)}}, 
        ymajorgrids=true,
        xmajorgrids=true,
        grid style=dashed,
        grid=both,
        grid style={solid, line width=.1pt, draw=gray!15},
        minor grid style={dotted, gray, line width=0.2pt},
    ]
    \addplot[
        color=black,
        line width=1pt,
        mark size=2pt,
        ]
    table[x=PhaseError,y=MSE_0dB]
    {images/Data/MSE_CSI.dat};
    \addplot[
        color=black,
        dash pattern=on 4pt off 2pt,
        line width=1pt,
        ]
    table[x=PhaseError,y=MSE_10dB]
    {images/Data/MSE_CSI.dat};
    \addplot[
        color=black,
        dash pattern=on 1pt off 1pt on 4pt off 2pt,
        line width=1pt,
        mark size=2pt,
        ]
    table[x=PhaseError ,y=MSE_20dB]
    {images/Data/MSE_CSI.dat};
     \addplot[dashed,
        color=black!30,
        dash pattern=on 1pt off 1pt,
        line width=1pt,
        mark size=2pt,
        ]
    table[x=PhaseError,y=MSE_30dB]
    {images/Data/MSE_CSI.dat};
    \legend{SNR (per user): $0$ dB, SNR (per user): $10$ dB, SNR (per user): $20$ dB, SNR (per user): $30$ dB};
    \end{axis}
\end{tikzpicture}
  \caption{
Impact of imperfect CSI on function computation as a function of maximum phase deviation.
}
  \label{fig:CSIerror}
\end{minipage}%
\hspace{20pt}
\begin{minipage}{.45\textwidth}
  \centering
     \begin{tikzpicture} 
    \begin{axis}[
        xlabel={$K$},
        ylabel={MSE},
        label style={font=\tiny},
        legend cell align={left},
        tick label style={font=\tiny} , 
        width=\textwidth,
        height=6cm,
        xmin=1, xmax=10,
        ymin=1e-2, ymax=3,
        xtick={1, 2, 3, 4, 5,6,7,8,9,10},
        ymode = log,
       legend style={nodes={scale=0.38, transform shape}, at={(0.98,0.38)}}, 
        ymajorgrids=true,
        xmajorgrids=true,
        grid style=dashed,
        grid=both,
        grid style={solid, line width=.1pt, draw=gray!25},
        minor grid style={dotted, gray, line width=0.2pt},
    ]
    \addplot[
        color=black,
        line width=1pt,
        mark size=2pt,
        ]
    table[x=K,y=MSE_SNR10_pha0]
    {images/Data/NMSEerrorphase.dat};
    \addplot[
        color=black,
        dash pattern=on 4pt off 2pt,
        line width=1pt,
        ]
    table[x=K,y=MSE_SNR10_pha30]
    {images/Data/NMSEerrorphase.dat};
    \addplot[
        color=black,
        dash pattern=on 1pt off 1pt on 4pt off 2pt,
        line width=1pt,
        mark size=2pt,
        ]
    table[x=K ,y=MSE_SNR10_pha60]
    {images/Data/NMSEerrorphase.dat};
     \addplot[dashed,
        color=black!30,
        dash pattern=on 1pt off 1pt,
        line width=1pt,
        mark size=2pt,
        ]
    table[x=K,y=MSE_SNR10_pha90]
    {images/Data/NMSEerrorphase.dat};
    \addplot[
        color=chestnut,
        line width=1pt,
        mark size=2pt,
        ]
    table[x=K,y=MSE_SNR20_pha0]
    {images/Data/NMSEerrorphase.dat};
    \addplot[
        color=chestnut,
        dash pattern=on 4pt off 2pt,
        line width=1pt,
        ]
    table[x=K,y=MSE_SNR20_pha30]
    {images/Data/NMSEerrorphase.dat};
    \addplot[
        color=chestnut,
        dash pattern=on 1pt off 1pt on 4pt off 2pt,
        line width=1pt,
        mark size=2pt,
        ]
    table[x=K ,y=MSE_SNR20_pha60]
    {images/Data/NMSEerrorphase.dat};
     \addplot[dashed,
        color=chestnut!30,
        dash pattern=on 1pt off 1pt,
        line width=1pt,
        mark size=2pt,
        ]
    table[x=K,y=MSE_SNR20_pha90]
    {images/Data/NMSEerrorphase.dat};
    \legend{
    {SNR (per user): $10$ dB, max. phase dev.: $0^{\circ}$},
    {SNR (per user): $10$ dB, max. phase dev.: $30^{\circ}$},
    {SNR (per user): $10$ dB, max. phase dev.: $60^{\circ}$},
    {SNR (per user): $10$ dB, max. phase dev.: $90^{\circ}$},
    {SNR (per user): $20$ dB, max. phase dev.: $0^{\circ}$},
    {SNR (per user): $20$ dB, max. phase dev.: $30^{\circ}$},
    {SNR (per user): $20$ dB, max. phase dev.: $60^{\circ}$},
    {SNR (per user): $20$ dB, max. phase dev.: $90^{\circ}$}};
    \end{axis}
\end{tikzpicture}
  \caption{
Impact of imperfect CSI on function computation as a function of number of users.}
  \label{fig:CSIerrorUsers}
\end{minipage}
\end{figure}

Finally, in \figurename~\ref{fig:CSIerror}, we demonstrate the impact of imperfect CSI on function computation, where the desired function is the sum of parameters. For this analysis, we consider $K = 10$ users, where each user chooses their parameters independently from a uniform distribution, with lower and upper limits $-\sqrt{3}$ and $\sqrt{3}$, respectively. Hence, the maximum and minimum values of the sum of parameters are $-10\sqrt{3}$ and $10\sqrt{3}$, respectively, and the variance of the parameters is 1. We model the imperfect CSI  by introducing phase synchronization errors to the transmitters. We model the phase error at each user with a uniform distribution and sweep the maximum phase deviation in this analysis. We design the precoders at the transmitters and receiver based on the discussions in Section~\ref{subsubsec:PowerControl} for a single-antenna scenario. In the simulations, we also consider the random realization of the fading channel and average out the impact of the channel and parameter realizations on MSE. The results in \figurename~\ref{fig:CSIerror} indicate that function computation is sensitive to imperfections and rapidly increases in the case of CSI errors. For example, if the phase synchronization error is $30^\circ$, the MSE increases by almost a factor of 10 for 30 dB SNR. In \figurename~\ref{fig:CSIerrorUsers}, we conduct the same analysis by sweeping the number of users. The curves show that the MSE increases with the number of users. Also, the imperfect CSI causes significant distortion in the computation.


\section{Security in AirComp}
\label{sec:security}


AirComp relies on the superposition of the transmitted signals. Hence, it has both positive and negative implications regarding security. On the one hand, the superposition in AirComp promotes user privacy as the transmitted signals cannot be directly observed. On the other hand, it opens up potential adversaries to harm the computation, and their impact cannot be directly inferred \cite{Poor, sahin23_survey}. The security aspects of AirComp may be investigated in four categories: Byzantine attacks, jamming, privacy, and eavesdropping.

Byzantine attacks are launched by the adversary or faulty nodes already part of the network, and AirComp is sensitive to these attacks. For example, for FEEL with AirComp, if the local data at the nodes are deliberately labeled incorrect (i.e., one of the data poisoning attacks) or the sign of the gradients are flipped (i.e., one of the model poisoning attacks) and the aggregation is handled through an AirComp scheme, the learning process can be unreliable. For AirComp, a Byzantine attack is a major problem because well-known precaution strategies relying on the observation of local information cannot be utilized directly as the computing node observes the superposed signal, not the signals transmitted themselves \cite{Fan_2022byzan}. To improve resilience against such attacks, several authors investigate the Weiszfeld algorithm and implement the corresponding iterations steps by considering AirComp, which is also in line with computing the median function or majority vote \cite{huang2021byzantineresilient}. Another strategy is that the network divides the nodes into multiple groups in orthogonal resources, uses AirComp for each group, and compares the distances among the groups \cite{park_2022tcom}.
Like Byzantine attacks, AirComp can be sensitive to jamming. Compared to the Byzantine attackers, jammers are not part of the network. Hence, it can be easier to detect the presence of jammers. One advantage of AirComp is that it is inherently distributed and can provide resilience against jammers, provided that the jamming signal is less powerful than the superposed signal. One way to improve resilience against jammers further is to use a common spreading code, assuming that the jammers do not know the spreading code. Hence, the interference due to the adversary is suppressed in the de-spreading operation at the computing node \cite{zhu20_baa}. 

AirComp promotes privacy as the local information is not observed. Also, it can accommodate existing privacy-enhancing methods. For example, a typical approach is to randomize the disclosed statistics by adding random or artificial noise, which causes a trade-off between accuracy and privacy \cite{Seif_2020isit}. For example, for FEEL, these random perturbations can be added to the model parameters or gradients before signal superposition to enhance privacy. Since AirComp can also be used with channel inversion at the transmitters, arguably, privacy can be promoted further \cite{Dongzhu_2021}.

AirComp can be sensitive to eavesdropping depending on the AirComp scheme. If the AirComp uses channel state information at the transmitters, an eavesdropper cannot directly observe the coherently superposed signal, and it provides a level of physical layer security against eavesdroppers. However, if AirComp relies on non-coherent computations for robust computation or does not rely on channels at the transmitters to address hardware impairments, it becomes vulnerable to eavesdropping. To address this issue, one approach is to divide the transmitter into two groups based on the amplitude of the channels and use a group of nodes with weaker channel conditions as jammers \cite{Hyoungsuk_2011}. Another method is to inject artificial noise at the computing node, where the computing node removes the noise from the signal superposition \cite{Frey_2021}. Also, homomorphic encryption methods can be used along with AirComp as they allow superposed symbols to be encrypted without the secret key \cite{Stoica_2022}.

\section{Conclusions and future research}

Future communication systems will focus on enabling higher-level tasks rather than merely transmitting data, necessitating redesigns to meet specific application or computational requirements. Reducing computational complexity and delays may become more critical than minimizing bit error rates. In this context, AirComp presents a highly attractive alternative. This article highlights the challenge AirComp poses for signal processing and waveform design in radio transmission. A systematic review of the different alternatives in the literature, together with the design guidelines to face the communication channel, are presented in this paper. 

The modulation sections expand the range of possible waveforms to include not only amplitude but also angular modulations, thanks to TBMA. Note that angular modulations are more attractive than amplitude ones due to their robustness against channel attenuation, as the information is not encoded in the waveform's amplitude. A general approach to the joint optimization of preprocessing, waveform, and postprocessing is also formulated in \eqref{eq:main_goal_fading}. One can think of future neural networks to address this complex joint optimization problem, where \gls{MSE} or other of the metrics that have been introduced in section \ref{sec:metrics} can be used according to the final application purposes. Note that these neural networks should consider the constraints imposed by the specific computation and application. These constraints are on: power, transmission bandwidth, number of nodes, quantization error, and maximum calculation error allowed. The complexity supported by the transmitting nodes must also be considered in this optimization. For example, due to the continuous demand for higher bandwidth in \gls{IoT} networks (e.g., an increased number of interconnected entities or multimedia sensors for surveillance), binary \gls{MSK} or differential modulations (e.g., differential \gls{PCM}) become appealing. They modulate the difference between successive samples rather than the sample itself, which helps reduce traffic with event-based sampling by selectively transmitting significant differences based on the application, resulting in an effective sampling rate below the Nyquist rate \cite{premaratne2021event}. Developing AirComp schemes for these \gls{IoT} modulation schemes is an open area of research.


Additionally, integrating compressed sensing, as discussed in \cite{gold1}, in the preprocessing stage presents an interesting challenge. With the increasing development of \gls{AI} systems and semantic communications, source encoders are expected to extract high-level features from data, providing gains similar to how traditional voice communication systems extract spectral content from voice data conveyed through the channel \cite{blau2018perception}.



Another critical aspect is designing the physical layer frame to introduce the most suitable multiplexing and/or diversity techniques based on the wireless channel and implementation constraints. To date, OFDM has emerged as the best solution for solving time synchronization problems between different transmitters and handling channels with frequency-selective fading. However, other non-multicarrier options may be more interesting when bandwidth is limited, as for instance, the MIMO techniques that we revisit in this paper. Note, however, that the existing signal processing techniques for \gls{AirComp} are mainly developed for \gls{DA}, but not in the context of \gls{TBMA}. Overall, many existing signal processing techniques designed to enhance the diversity or multiplexing capabilities of point-to-point or non-AirComp MAC systems can potentially be adapted and applied to AirComp. { 
AirComp can benefit from channel manipulation techniques like intelligent surfaces \cite{An_RIS2024}. However, we should note these approaches require more justification in terms of signal processing considering practical conditions like mobility and validations beyond theoretical analysis. These open up a whole range of problems within signal processing.

Finally, the design of channel codes for AirComp is an open research topic. AirComp requires new channel codes that comply with the additive nature of the scheme, meaning the addition of redundancy must provide useful information at the receiver \cite{xie2023joint}. Some ideas may be borrowed from coded computing \cite{Li2020Coded}, which aims to overcome fundamental bottlenecks in wireless systems with massive numbers of distributed computing nodes, such as communication complexity and latency. As several challenges remain unresolved when underlying codes are designed over discrete spaces (e.g., overflow errors), analog error-correcting codes 
are suitable for applications, such as learning, that are insensitive to controlled inaccuracies. Methods and ideas for optimal codes over discrete spaces cannot be extended to continuous spaces due to fundamental differences. These issues have motivated researchers to rethink code design over continuous spaces for specific applications where data is inherently real/complex-valued \cite{soleymani2022codes}. AirComp can benefit from and follow these new coding trends, presenting a green field for exploration with signal processing.

\bibliographystyle{IEEEtran}
\bibliography{references.bib}

\begin{thebibliography}{10}
\providecommand{\url}[1]{#1}
\csname url@samestyle\endcsname
\providecommand{\newblock}{\relax}
\providecommand{\bibinfo}[2]{#2}
\providecommand{\BIBentrySTDinterwordspacing}{\spaceskip=0pt\relax}
\providecommand{\BIBentryALTinterwordstretchfactor}{4}
\providecommand{\BIBentryALTinterwordspacing}{\spaceskip=\fontdimen2\font plus
\BIBentryALTinterwordstretchfactor\fontdimen3\font minus \fontdimen4\font\relax}
\providecommand{\BIBforeignlanguage}[2]{{%
\expandafter\ifx\csname l@#1\endcsname\relax
\typeout{** WARNING: IEEEtran.bst: No hyphenation pattern has been}%
\typeout{** loaded for the language `#1'. Using the pattern for}%
\typeout{** the default language instead.}%
\else
\language=\csname l@#1\endcsname
\fi
#2}}
\providecommand{\BIBdecl}{\relax}
\BIBdecl

\bibitem{goldenbaum13_aircomp}
M.~Goldenbaum, H.~Boche, and S.~Stanczak, ``Harnessing interference for analog function computation in wireless sensor networks,'' \emph{IEEE Transactions on Signal Processing}, vol.~61, no.~20, pp. 4893--4906, 2013.

\bibitem{nazer07_aircomp}
B.~Nazer and M.~Gastpar, ``Computation over multiple-access channels,'' \emph{IEEE Transactions on Information Theory}, vol.~53, no.~10, pp. 3498--3516, 2007.

\bibitem{gastpar06_liaison}
M.~Gastpar, M.~Vetterli, and P.~Dragotti, ``Sensing reality and communicating bits: a dangerous liaison,'' \emph{IEEE Signal Processing Magazine}, vol.~23, no.~4, pp. 70--83, 2006.

\bibitem{Gunduz}
D.~Gunduz, Z.~Qin, I.~Estella, H.~S. Dhillon, Z.~Yang, A.~Yener, K.~K. Wong, and C.-B. Chae, ``Beyond transmitting bits: Context, semantics, and task-oriented communications,'' \emph{IEEE Journal on Selected Areas in Communications}, vol.~41, no.~1, pp. 5--41, 2023.

\bibitem{Poor}
H.~Hellström, J.~M.~B. da~Silva~Jr., M.~M. Amiri, M.~Chen, V.~Fodor, H.~V. Poor, and C.~Fischione, \emph{Wireless for Machine Learning: A Survey}.\hskip 1em plus 0.5em minus 0.4em\relax Foundations and Trends® in Signal Processing, 2022, vol.~15, no.~4.

\bibitem{islam2019noma}
S.~Islam, M.~Zeng, O.~A. Dobre, and K.-S. Kwak, ``Non-orthogonal multiple access (noma): How it meets 5g and beyond,'' \emph{arXiv preprint arXiv:1907.10001}, 2019.

\bibitem{sahin23_survey}
A.~\c{S}ahin and R.~Yang, ``A survey on over-the-air computation,'' \emph{IEEE Communications Surveys \& Tutorials}, vol.~25, no.~3, pp. 1877--1908, 2023.

\bibitem{wang22_foundations}
Z.~Wang, Y.~Zhao, Y.~Zhou, Y.~Shi, C.~Jiang, and K.~B. Letaief, ``Over-the-air computation: Foundations, technologies, and applications,'' \emph{arXiv preprint arXiv:2210.10524}, 2022.

\bibitem{mish}
M.~Dohler, R.~Heath, A.~Lozano, C.~Papadias, and R.~Valenzuela, ``Is the phy layer dead?'' \emph{IEEE Communications Magazine}, vol.~49, no.~4, pp. 159--165, 2011.

\bibitem{liu2024recent}
B.~Liu, N.~Lv, Y.~Guo, and Y.~Li, ``Recent advances on federated learning: A systematic survey,'' \emph{Neurocomputing}, p. 128019, 2024.

\bibitem{scholkopf2007mapreduce}
C.-T. Chu, S.~Kim, Y.-A. Lin, Y.~Yu, G.~Bradski, K.~Olukotun, and A.~Ng, ``Map-reduce for machine learning on multicore,'' \emph{Advances in neural information processing systems}, vol.~19, pp. 281--288, 2006.

\bibitem{wu16_stac}
X.~Wu, S.~Zhang, and A.~Ozgur, ``Stac: Simultaneous transmitting and air computing in wireless data center networks,'' \emph{IEEE Journal on Selected Areas in Communications}, vol.~34, no.~12, pp. 4024--4034, 2016.

\bibitem{song2011camera}
B.~Song, C.~Ding, A.~T. Kamal, J.~A. Farrell, and A.~K. Roy-chowdhury, ``Distributed camera networks,'' \emph{IEEE Signal Processing Magazine}, vol.~28, no.~3, pp. 20--31, 2011.

\bibitem{liu2023pooling}
Z.~Liu, Q.~Lan, A.~E. Kalør, P.~Popovski, and K.~Huang, ``Over-the-air view-pooling for low-latency distributed sensing,'' in \emph{2023 IEEE 24th International Workshop on Signal Processing Advances in Wireless Communications (SPAWC)}, 2023, pp. 71--75.

\bibitem{lee2023platoon}
J.~Lee, Y.~Jang, H.~Kim, S.-L. Kim, and S.-W. Ko, ``Over-the-air consensus for distributed vehicle platooning control,'' in \emph{ICC 2023 - IEEE International Conference on Communications}, 2023, pp. 5965--5971.

\bibitem{sahin23_kmeans}
A.~\c{S}ahin, ``Wireless federated $k$-means clustering with non-coherent over-the-air computation,'' in \emph{Proc. IEEE Military Communications Conference (MILCOM)}, 2023, pp. 339--344.

\bibitem{guirado23whype}
R.~Guirado, A.~Rahimi, G.~Karunaratne, E.~Alarcon, A.~Sebastian, and S.~Abadal, ``Whype: A scale-out architecture with wireless over-the-air majority for scalable in-memory hyperdimensional computing,'' \emph{IEEE Journal on Emerging and Selected Topics in Circuits and Systems}, vol.~13, no.~1, pp. 137--149, 2023.

\bibitem{gold1}
J.-J. Xiao, S.~Cui, Z.-Q. Luo, and A.~J. Goldsmith, ``Linear coherent decentralized estimation,'' \emph{IEEE Transactions on Signal Processing}, vol.~56, no.~2, pp. 757--770, 2008.

\bibitem{goldenbaum15_lattice}
M.~Goldenbaum, H.~Boche, and S.~Stanczak, ``Nomographic functions: Efficient computation in clustered gaussian sensor networks,'' \emph{IEEE Transactions on Wireless Communications}, vol.~14, no.~4, pp. 2093--2105, 2015.

\bibitem{martinez23_enn}
M.~Martinez-Gost, A.~P{\'e}rez-Neira, and M.~{\'A}. Lagunas, ``E{N}{N}: A neural network with {D}{C}{T}-adaptive activation functions,'' \emph{IEEE Journal of Selected Topics in Signal Processing}, 2023.

\bibitem{carlson}
A.~B. Carlson and P.~Crilly, \emph{Communication systems}.\hskip 1em plus 0.5em minus 0.4em\relax McGraw-Hill Education, 2009.

\bibitem{zhu20_baa}
G.~Zhu, Y.~Wang, and K.~Huang, ``Broadband analog aggregation for low-latency federated edge learning,'' \emph{IEEE Transactions on Wireless Communications}, vol.~19, no.~1, pp. 491--506, 2020.

\bibitem{goldenbaum13_robust}
M.~Goldenbaum and S.~Stanczak, ``Robust analog function computation via wireless multiple-access channels,'' \emph{IEEE Transactions on Communications}, vol.~61, no.~9, pp. 3863--3877, 2013.

\bibitem{cao20_powercontrol}
X.~Cao, G.~Zhu, J.~Xu, and K.~Huang, ``Optimized power control for over-the-air computation in fading channels,'' \emph{IEEE Transactions on Wireless Communications}, vol.~19, no.~11, pp. 7498--7513, 2020.

\bibitem{premaratne2021event}
U.~Premaratne, S.~Warnakulasooriya, and R.~Nandana, ``Characterization of event-based sampling encoders for industrial internet of things using input–output mutual information,'' \emph{IEEE Transactions on Industrial Informatics}, vol.~17, no.~8, pp. 5495--5505, 2021.

\bibitem{xie2023joint}
X.~Xie, C.~Hua, J.~Hong, and Y.~Wei, ``Joint design of coding and modulation for digital over-the-air computation,'' \emph{arXiv preprint arXiv:2311.06829}, 2023.

\bibitem{mergen06_tbma}
G.~Mergen and L.~Tong, ``Type based estimation over multiaccess channels,'' \emph{IEEE Transactions on Signal Processing}, vol.~54, no.~2, pp. 613--626, 2006.

\bibitem{martinez23_tbma}
M.~Martinez-Gost, A.~Pérez-Neira, and M.~{\'A}. Lagunas, ``{LoRa}-based over-the-air computing for {Sat-IoT},'' in \emph{2023 31st European Signal Processing Conference (EUSIPCO)}, 2023, pp. 1514--1518.

\bibitem{sahin22_numerals}
A.~\c{S}ahin, ``Over-the-air computation based on balanced number systems for federated edge learning,'' \emph{IEEE Transactions on Wireless Communications}, vol.~23, no.~5, pp. 4564--4579, 2024.

\bibitem{martinez2024log}
M.~Martinez-Gost, A.~P{\'e}rez-Neira, and M.~{\'A}. Lagunas, ``Log-{F}{S}{K}: A frequency modulation for over-the-air computing,'' \emph{EUSIPCO}, 2024.

\bibitem{gost23_dct}
M.~M. Gost, A.~Pérez-Neira, and M.~A. Lagunas, ``{DCT}-based air interface design for function computation,'' \emph{IEEE Open Journal of Signal Processing}, vol.~4, pp. 44--51, 2023.

\bibitem{guirado2023whype}
R.~Guirado, A.~Rahimi, G.~Karunaratne, E.~Alarc{\'o}n, A.~Sebastian, and S.~Abadal, ``Whype: A scale-out architecture with wireless over-the-air majority for scalable in-memory hyperdimensional computing,'' \emph{IEEE Journal on Emerging and Selected Topics in Circuits and Systems}, vol.~13, no.~1, pp. 137--149, 2023.

\bibitem{razavikia23_channelcomp}
S.~Razavikia, J.~M.~B. Da~Silva, and C.~Fischione, ``Channelcomp: A general method for computation by communications,'' \emph{IEEE Transactions on Communications}, pp. 1--1, 2023.

\bibitem{Sidir2006Physical}
N.~Sidiropoulos, T.~Davidson, and Z.-Q. Luo, ``Transmit beamforming for physical-layer multicasting,'' \emph{IEEE Transaction on Signal Processing}, vol.~54, no.~6, pp. 2239--2251, 2006.

\bibitem{vandenberghe1996semidefinite}
L.~Vandenberghe and S.~Boyd, ``Semidefinite programming,'' \emph{SIAM review}, vol.~38, no.~1, pp. 49--95, 1996.

\bibitem{luo2010semidefinite}
Z.-Q. Luo, W.-K. Ma, A.~M.-C. So, Y.~Ye, and S.~Zhang, ``Semidefinite relaxation of quadratic optimization problems,'' \emph{IEEE Signal Processing Magazine}, vol.~27, no.~3, pp. 20--34, 2010.

\bibitem{razavikia2023sumcomp}
S.~Razavikia, J.~M. B. D.~S. Júnior, and C.~Fischione, ``Sumcomp: Coding for digital over-the-air computation via the ring of integers,'' 2023.

\bibitem{sahin23_noncoherent}
A.~\c{S}ahin, ``Distributed learning over a wireless network with non-coherent majority vote computation,'' \emph{IEEE Transactions on Wireless Communications}, vol.~22, no.~11, pp. 8020--8034, 2023.

\bibitem{martinez23_feel}
M.~Martinez-Gost, A.~Perez-Neira, and M.~Ã. Lagunas, ``Frequency modulation aggregation for federated learning,'' in \emph{GLOBECOM 2023 - 2023 IEEE Global Communications Conference}, 2023, pp. 1878--1883.

\bibitem{zhu21_obda}
G.~Zhu, Y.~Du, D.~Gunduz, and K.~Huang, ``One-bit over-the-air aggregation for communication-efficient federated edge learning: Design and convergence analysis,'' \emph{IEEE Transactions on Wireless Communications}, vol.~20, no.~3, pp. 2120--2135, 2021.

\bibitem{zhu19_mimo}
G.~Zhu and K.~Huang, ``Mimo over-the-air computation for high-mobility multimodal sensing,'' \emph{IEEE Internet of Things Journal}, vol.~6, no.~4, pp. 6089--6103, 2019.

\bibitem{Fan_2022byzan}
X.~Fan, Y.~Wang, Y.~Huo, and Z.~Tian, ``{BEV-SGD}: Best effort voting {SGD} against byzantine attacks for analog aggregation based federated learning over the air,'' \emph{IEEE Internet of Things Journal}, pp. 1--14, 2022.

\bibitem{huang2021byzantineresilient}
S.~Huang, Y.~Zhou, T.~Wang, and Y.~Shi, ``Byzantine-resilient federated machine learning via over-the-air computation,'' in \emph{Proc. IEEE International Conference on Communications Workshops (ICC Workshops)}, 2021, pp. 1--6.

\bibitem{park_2022tcom}
S.~Park and W.~Choi, ``Byzantine fault tolerant distributed stochastic gradient descent based on over-the-air computation,'' \emph{IEEE Trans. Commun.}, pp. 1--15, 2022.

\bibitem{Seif_2020isit}
M.~Seif, R.~Tandon, and M.~Li, ``Wireless federated learning with local differential privacy,'' in \emph{Proc. IEEE International Symposium on Information Theory (ISIT)}, 2020, pp. 2604--2609.

\bibitem{Dongzhu_2021}
D.~Liu and O.~Simeone, ``Privacy for free: Wireless federated learning via uncoded transmission with adaptive power control,'' \emph{IEEE Journal on Selected Areas in Communications}, vol.~39, no.~1, pp. 170--185, 2021.

\bibitem{Hyoungsuk_2011}
H.~Jeon, D.~Hwang, J.~Choi, H.~Lee, and J.~Ha, ``Secure type-based multiple access,'' \emph{IEEE Transactions on Information Forensics and Security}, vol.~6, no.~3, pp. 763--774, 2011.

\bibitem{Frey_2021}
M.~Frey, I.~Bjelakovic, and S.~Stanczak, ``Towards secure over-the-air computation,'' in \emph{Proc. IEEE International Symposium on Information Theory (ISIT)}, 2021, pp. 700--705.

\bibitem{Stoica_2022}
R.-A. Stoica, O.~Taghizadeh, and S.~B. Mary~Baskaran, ``Secret computing over multiple access channels,'' in \emph{2022 56th Asilomar Conference on Signals, Systems, and Computers}, 2022, pp. 559--563.

\bibitem{blau2018perception}
Y.~Blau and T.~Michaeli, ``The perception-distortion tradeoff,'' in \emph{Proceedings of the IEEE conference on computer vision and pattern recognition}, 2018, pp. 6228--6237.

\bibitem{An_RIS2024}
J.~An, C.~Yuen, C.~Xu, H.~Li, D.~W.~K. Ng, M.~Di~Renzo, M.~Debbah, and L.~Hanzo, ``Stacked intelligent metasurface-aided mimo transceiver design,'' \emph{IEEE Wireless Communications}, vol.~31, no.~4, pp. 123--131, 2024.

\bibitem{Li2020Coded}
S.~Li and S.~Avestimehr, ``Coded computing: Mitigating fundamental bottlenecks in large-scale distributed computing and machine learning,'' \emph{Foundations and Trends in Communications and Information Theory}, vol.~17, no.~1, pp. 1--148, 2020.

\bibitem{soleymani2022codes}
M.~Soleymani, ``Analog coding: Theory and applications,'' PhD thesis, University of Michigan, Michigan, MI, 2022.

\end{thebibliography}

\end{document}